\documentstyle[11pt,epsfig,aaspp]{article}
\begin{document}

\textwidth 6.5in
\textheight 8.5in
\topmargin 0.0in
\oddsidemargin 0.0in

\title{Lunar Outgassing, Transient Phenomena and The Return to The Moon\\
I: Existing Data}

\author{Arlin P.S.~Crotts}
\affil{Department of Astronomy, Columbia University,
Columbia Astrophysics Laboratory,\\
550 West 120th Street, New York, NY 10027}

\begin{abstract}
Transient lunar phenomena (TLPs) have been reported for centuries, but their
nature is largely unsettled, and remains controversial.
In this Paper I the database of TLP reports is subjected to a discriminating
statistical filter robust against sites of spurious reports, and produces a
restricted sample that may be largely reliable.
This subset is highly correlated geographically with the catalog of outgassing
events seen by the {\it Apollo 15, 16} and {\it Lunar Prospector}
alpha-particle spectrometers for episodic $^{222}$Rn gas release.

Both this robust TLP sample and even the larger, unfiltered sample are highly
correlated with the boundary between mare and highlands, as are both deep and
shallow moonquakes, as well as $^{220}$Po, a long-lived product of $^{222}$Rn
decay and a further tracer of outgassing.
This offers a second significant correlation relating TLPs and outgassing, and
may tie some of this activity to sagging mare basalt plains (perhaps mascons).

Additionally, low-level but likely significant TLP activity is connected to
recent, major impact craters (while moonquakes are not), which may indicate the
effects of cracks caused by the impacts, or perhaps avalanches, allowing
release of gas.
The majority of TLP (and $^{222}$Rn) activity, however, is confined to one site
that produced much of the basalt in the Procellarum Terrane.
Since this Terrane is antipodal to the huge South Pole-Aitken impact, it seems
plausible that this TLP activity may be tied to residual outgassing from the
formerly largest volcanic effusion sites from the deep lunar interior.

In Paper II of this series we discuss likely theoretical implications of past
and present outgassing and its connection to TLPs.
Paper III presents methodologies for remote and in-situ observations of TLPs
and manifestations of lunar outgassing.
These include several ground-based methods which can be implemented immediately.

With the coming exploration in the next few years by a fleet of robotic
spacecraft, followed by human exploration, the study of TLPs and outgassing is
both promising and imperiled.
We will have an unprecedented opportunity to study lunar outgassing, but will
also deal with a greater burden of anthropogenic lunar gas than ever produced,
probably more than the natural atmosphere itself.
There is a pressing need to study the lunar atmosphere and its potential
sources while it is still in its pristine state.
\end{abstract}

\medskip
\section{Introduction}

In the minds of many scientists, the Moon is a dead world.
Indeed, the Moon shows little activity compared to many bodies of its size or
larger.
Internal movements tend to be very low amplitude (see Nakamura et al.\ 1981,
for example) and the native atmosphere is typically at total atomic/molecular
number density below $10^5$ cm$^{-3}$ at the lunar surface (Hodges 1975,
Hoffman \& Hodges 1974) with a mass of 20-30 tonnes.
Some geological features are suggestive of recent activity (e.g., Schultz,
Staid \& Pieters 2001, 2006; c.f., Strain \& El-Baz 1980), but not overwhelming
in number.
As little as 3~Gy ago, however, the Moon has suffered a major fraction of its
surface covered by a high-temperature, refractory basalt.
Cooling models predict that the Moon has evolved to have a lithosphere of
essentially a single crustal plate many hundreds of km thick (e.g., Spohn 2004
\& op cit.); however, it is natural to wonder what evidence might exist for
residual volcanic activity persisting to the present.
This might be manifest in the form of volatile release to the surface through
partial breaching of the crust's integrity in the form of lithospheric
fracturing due to massive impacts, or stresses from tides or mascons
interacting with overlying crust (Reindler \& Arkani-Hame 2001).
In this paper I consider indications of rapid changes that may occur on the
Moon due to internal or intrinsic processes.
In accompanying papers we propose how to advance our understanding of this
situation beyond its current ambiguity.

Transient Lunar Phenomena (TLPs, called LTPs by some authors) are predominantly
optical-wavelength effects, typically reported by observers monitoring the Moon
visually through a telescope (with a few exceptions discussed below).
Their physical nature is unknown and even their reality is a point of dispute.
(I discuss in \S 3 episodic lunar gas discharges that I do not call TLPs {\it
per se} -- while some authors do -- but their relation to TLPs is a crucial
theme of this paper.)~
TLPs are usually brightenings or dimming/obscurations of the lunar surface,
sometimes changing color - usually reddish but occasionally blue or violet.
(Some early observers refer enigmatically to TLPs only as lunar ``volcanoes''.)~
TLPs are localized, nearly always to a radius much less than 100 km, often as
unresolved points (less than an arcsecond - about 1.9 km at the Earth-Moon
distance).
There are classes of phenomena, however, that some authors call TLPs that
involve the whole Moon and, while interesting, will fall outside our discussion
(some examples: Spinrad 1964, Sanduleak \& Stocke 1965, Verani et al.\ 2001).
I do not discuss phenomena tied to solar eclipses (but retain a few during
lunar eclipses), and will omit events where the particular location is
unspecified, including several events involving the extension of the cusps of
the crescent Moon.
TLPs are reported on timescales from ``instantaneous'' (probably a few tenths
of a second due to small meteorite impacts) to several hours.
TLPs are reported for many sites over the lunar surface, but are far from
randomly distributed; a key question is whether this is physical or a severe
observer bias.

Even casual investigators of TLPs notice something unusual associated
with the region around the crater Aristarchus.
This includes the adjacent crater Herodotus, and Schr\"oter's Valley (Vallis
Schr\"oteri) flowing from ``Cobra-Head'' (or ``Cobra's Head''), which together
occupy the southeastern quarter of the compact, raised Aristarchus Plateau of
$\sim$40,000 km$^2$ within the huge ($4\times 10^6$ km$^2$) mare region Oceanus
Procellarum, but close to the Mare Imbrium boundary.
(Vallis Schr\"oteri was once selected as the landing site for Apollo 18, later
cancelled along with Apollos 19 and 20.)~
Aristarchus is among the brightest nearside lunar craters, sometimes {\it the}
brightest, sometimes visible to the unaided eye from Earth\footnote{One even
finds apparent reference to Aristarchus in Tang Dynasty (618 -- 907 A.D./CE)
writings (Mayers 1874).} along with perhaps Copernicus and Tycho (which each
produce less than 5\% of the TLP reports of Aristarchus), and one of the
youngest.

More than Copernicus, Aristarchus is distinguished by its stark contrast to the
surrounding dark background (but this is unlike other TLP-producing features on
the Plateau).
Once the region was intensely active, with volcanic flows and eruptions, and
many sinuous rilles remain, likely old lava channels, including the most
voluminous on the Moon, Vallis Schr\"oteri.
Not only is this region responsible for $\sim$50\% of the visual
telescopic TLP reports (but also likely receives a disproportionate fraction of
the observing attention), undeniably objective lunar anomalies of a transient,
physical nature occur in the Aristarchus region, as detailed below.

%

Several experiments on Apollo lunar missions, orbiting and surface, as well as
on {\it Lunar Prospector}, were designed to detect and identify gases in the
tenuous lunar atmosphere, both ions and neutral species, plus decay products
from gaseous radioactive isotopes.
Even though some of these spent only days or weeks operating near the Moon,
most observed evidence of sporadic outgassing activity, including events that
seem unassociated with anthropogenic effects.
This paper treats the correspondence between this activity and TLPs.
To establish if TLP behavior is connected with the physics of the lunar
environment, in Paper II we explore ways in which this might be so, and in
Paper III, I explore ways in which this understanding can be increased with
technologically-accessible, systematic observations.

In the next decade, numerous spacecraft and humans will visit the Moon again.
This offers an unprecedented opportunity to study the atmosphere of the Moon,
but will also introduce transients from human activity that may complicate our
understanding of this gas and what it can disclose regarding the lunar
interior's structure, composition and evolution.
We must evaluate the current results now and expand upon them rapidly to
exploit our upcoming opportunity to explore the Moon in its still pristine
state.

\section{Transient Lunar Phenomena}

TLP as observed are apparently rare events, and therefore the TLP database is
largely anecdotal.
Furthermore, since TLPs are observed for short durations, there is rarely the
opportunity to introduce possibly corroborating observations, such as
photography or spectroscopy.
For these reasons, primarily, the reputation of TLPs among many scientists is
suspect, and also their explanation is largely unsettled.
Nonetheless, TLPs represent a large database, cataloged by the great efforts of
Winfred Cameron, Barbara Middlehurst and others, and if the reality of TLPs can
be evaluated they might offer a potentially interesting method to study the
Moon.

In the companion to this work (Paper 0), the statistical structure of the TLP
database is investigated in terms of the sensitivity of the results (that is,
the consistent rate at which specific lunar features produce TLP reports) to
parameters that might betray human observer bias or error.
Paper 0 concludes that to the extent that one can test the
effects of human bias/errors, they appear important only during one historical
period (the most recent), and otherwise the behavior of TLP source features are
impressively consistent, qualitatively.
These tests imply $\ga 10$\% of TLP reports are inconsistent and therefore
suspect, but that many, quite possibly the majority, indicate a consistent
phenomenon untied to the non-lunar variables involving the manner of
observation.

Below I will review the arguments for statistical consistency from Paper 0 by
considering such a ``robustness test,'' but first deal with a few auxiliary
issues.
There are several works which have studied TLPs in terms of the fraction of
various kinds of events e.g., brightenings, dimmings, obscurations, red and
blue-colored events.
Classification and analysis of TLP categories have been discussed at length in
the literature; refer to Cameron (1972) for an effective overview.
Of 113 reports in Middlehurst (1977a) involving enhanced brightness in blue
and/or violet, 101 of them involve J.C.\ Bartlett, composing most of his total
of 114 reports (between 1949 and 1966), most of those (108) involving
Aristarchus.
In contrast only 9 of 12 total non-Bartlett blue/violet events occur in the
same years (during which 47\% of all reports occur).
We must correct for this somehow, either by rejecting all blue/violet events or
all reports by Bartlett; I choose the latter.

A total of 71 reports in Middlehurst et al.\ (1968) include duration estimates
(which can be interpreted to better than a factor of 2).
Of course this is not a statistical sample, but the reports indicate prolonged
occurrences; binned in $\sqrt{10}$ intervals from 60s to 19000s (with the
longest event being 18000s and the shortest 60s) the duration distribution is:
60-190s, 7 reports; 191-600s: 9; 601-1900s, 27, 1901-6000s, 23; more than
6000s: 5.
These effects are not so rapid as to not allow reinspection (albeit by the
same observer in most cases).
There are four cases in Middlehurst et al.\ (1968) described as sudden,
isolated flashes of light, and these are not correlated with meteor showers
(the TLPs occurring on 1945 October 19, 1955 April 24, 1957 October 12, and
1967 September 11).
None of these are well-placed with respect to known meteor showers.
(April 23 is the peak of the Pi Puppids, but these are strong only near the
perihelion of comet 26P/Grigg-Skjellerup, which occurred in 1952 and 1957, not
1955.)~ 
Suggestions for other mechanisms for short-lived TLPs include piezoelectric
discharge (Kolovos et al.\ 1988, 1992 - which also includes an interesting
recorded TLP observation).
In Paper 0 we show that evidence indicates that minimal impacts visible from
Earth will occur on sub-second timescales, while even the brightest and rarest
impacts will be visible for only a few seconds.

Even if very large impacts can produce events of sufficiently long duration, it
is clear from model computation e.g., Morrison et al.\ (1993) that the fresh
impacts seen in {\it Clementine} and other data sets cannot sustain such
activity.
This indicates that the great majority of TLP reports cannot be generated by
impacts, which leaves open the possibility that their geographical distribution
is not random.
The spatial distribution might be expected to carry detailed information about
the TLP mechanisms, assuming observer selection effects can be removed.
This will be the topic of some discussion below, but first note the results
from the raw catalogs.
Table 1 and Figure 1 are derived from reports listed by Middlehurst et
al.\ (1968), sometimes with additional information (but not additional reports)
drawn from Cameron (1978).

\noindent
Table 1: Number of TLPs Reported, by Feature
\begin{verbatim}
____________________________________________________________________________
Feature                 Lat Long  #      Feature                 Lat Long  # 
----------------------- --- --- ---      ----------------------- --- --- --- 
Aristarchus             24N 48W 122      Arzachel                18S  2W   1 
Plato                   51N  9W  40      Birt                    22S  9W   1 
Vallis Schroteri        26N 52W  20      Carlini                 34N 24W   1 
Alphonsus               13S  3W  18      Cavendish               24S 54W   1 
Gassendi                18S 40W  16      Censorinus               0N 32E   1 
Ross D                  12N 22E  13      Clavius                 58S 14W   1 
Mare Crisium            18N 58E  12      Conon                   22N  2E   1 
Cobra Head              24N 48W   6      Daniell                 35N 31E   1 
Copernicus              10N 20W   6      Darwin                  20S 69W   1 
Kepler                   8N 38W   6      Dawes                   17N 26E   1 
Posidonius              32N 30E   6      Dionysius                3N 17E   1 
Tycho                   43S 11W   6      Endymion                54N 56E   1 
Eratosthenes            15N 11W   5      Fracastorius            21S 33E   1 
Messier                  2N 48E   5      Godin                    2N 10E   1 
Grimaldi                 6S 68W   4      Hansteen                11S 52W   1 
Lichtenberg             32N 68W   4      Hercules                47N 39E   1 
Mons Piton              41N  1W   4      Herschel                 6S  2W   1 
Picard                  15N 55E   4      Humboldt                27S 80E   1 
Capuanus                34S 26W   3      Hyginus N                8N  6E   1 
Cassini                 40N  5E   3      Kant                    11S 20E   1 
Eudoxus                 44N 16E   3      Kunowsky                 3N 32W   1 
Mons Pico               46N  9W   3      Lambert                 26N 21W   1 
Pitatus                 30S 13W   3      Langrenus                9S 61E   1 
Proclus                 16N 47E   3      Leibnitz Mt. (unoffic.) 83S 39W   1 
Ptolemaeus               9S  2W   3      Manilius                15N  9E   1 
Riccioli                 3S 74W   3      Mare Humorum            24S 39W   1 
Schickard               44S 26E   3      Mare Nubium             10S 15W   1 
Theophilus              12S 26E   3      Mare Serenitatis        28N 18E   1 
Plato, 1.3' S.E. of     47N  3W   2      Mare Vaporum            13N  3E   1 
Alpetragius             16S  5W   2      Marius                  12N 51W   1 
Atlas                   47N 44E   2      Menelaus                16N 16E   1 
Bessel                  22N 18E   2      Mersenius               22S 49W   1 
Calippus                39N 11E   2      Mont Blanc              45N  0E   1 
Helicon                 40N 23W   2      Montes Carpatus         15N 25W   1 
Herodotus               23N 50W   2      Montes Taurus           26N 36E   1
Littrow                 21N 31E   2      Peirce A                18N 53E   1 
Macrobius               21N 46E   2      Philolaus               72N 32W   1 
Mare Humorum            24S 39W   2      Plinius                 15N 24E   1 
Mare Tranquilitaties     8N 28E   2      Sabine                   1N 20E   1 
Mons La Hire            28N 26W   2      Sinus Iridum, S. of     45N 32W   1 
Montes Alps, S. of      46N  2E   2      Sulpicius Gallus        20N 12E   1 
Montes Teneriffe        47N 13W   2      Taruntius                6N 46E   1 
Pallas                   5N  2W   2      Thales                  62N 50E   1 
Promontorium Agarum     18N 58E   2      Triesnecker              4N  4E   1 
Promontorium Heraclides 14N 66E   2      Vitruvius               18N 31E   1 
South Pole              90S  0E   2      Walter                  33S  0E   1 
Theaetetus              37N  6E   2                                          
Timocharis              27N 13W   2      ___________________________________  
Agrippa                  4N 11E   1      (unknown)               --- ---  43 
Anaximander             67N 51W   1      (cusps)                 --- ---  14 
Archimedes              30N  4W   1      (global)                --- ---   4 
____________________________________________________________________________
\end{verbatim}

There is a tendency for TLP reports to favor the western half of the near side
(106 in the east, 166 in the west in addition to 144 on the Aristarchus
plateau), which runs counter to the usual preference of casual observers to
observe earlier in the night.
This may be due to the greater extent of maria (and mare boundaries: see
the following) on the western side.
The primary spatial modulation of the report rate, that has been noticed
previously, beyond just the frequency at specific sites is the tendency of
reports to avoid the deep highlands and to some degree the mid-mare plains, but
instead to congregate in the vicinity of the maria/highland interface (Cameron
1967, 1972, Middlehurst \& Moore 1967, Buratti et al.\ 2000).
Even Aristarchus/Vallis Schr\"oteri/Cobra's Head/Herodotus in the midst of
Oceanus Procellarum rests on a giant block of about 40,000 km$^2$ (probably
from a previous mare basin impact) elevated 2 km above the mare plain, although
this might easily be a special case.

How do we deal statistically with the horrendous selection effects introduced
into this data set by the patterns and biases of the observers, most of whom
never intended that their reports form part of a statistical database?
This is as much a historical and even a psychological question as a
physical/mathematical one; however, there are some regularities that we might
exploit.
First, the pattern of TLP observer behavior seems to have changed significantly
in the mid-20th century, when well-publicized reports such as Alter (1957) drew
attention to TLPs and particular locations such as Alphonsus and Aristarchus.
Many observers after that era concentrated specifically on sites such as these
in an effort to maximize success in detecting a TLP.
Prior to this era, I see little evidence (Paper 0) that observers were drawn
{\it a priori} nearly as much to specific sites.
Middlehurst (1977a) has reviewed historical reports extensively and comes to a
similar conclusion.
Indeed, many reports from previous centuries neglect to fully specify the site
of their TLP.

I cannot fully appreciate the observing motivations of astronomers from so long
ago, but there is little written indicating special sites such as Aristarchus as
targets of propagating popular or professional attention in terms of TLPs
(Paper 0).
Aristarchus did receive wider scrutiny in 1911 when R.\ Wood indicated that it
might contain high concentrations of sulfur, but this did not produce a spate
of Aristarchus TLP reports.
Indeed, Wood discusses volcanism in the context of Aristarchus (sometimes known
as ``Wood's Spot''\footnote{e.g., Whitaker (1972), or
http://www.lpod.org/archive/archive/2004/01/LPOD-2004-01-17.htm}) and seems
unaware of the number of TLP reports in the vicinity (Wood 1911).
In Paper 0 earlier works by W.H.\ Pickering (1892, 2004) on Aristarchus and
lunar activity are detailed, but these show no evidence of having inspired
later TLP reports.
Furthermore, Birt (1870) and Whitley (1870) provide a historical overview
(1787--1880) of visual observations of Aristarchus (and Herodotus) while
conducting a spirited debate about the nature of features including possible
changes in their appearance.
They mention small, possible changes, but give them no special
significance, nor mention anything that today we might refer to as a
recognized TLP phenomenon (or at least a human tendency to report TLPs).
A different statement is made by Elger (1884), who again reviews Aristarchus,
Herodotus and the surrounding plateau.
While he does not mention anything like TLPs, he makes a telling statement:
``Although no part of the moon's visible surface has been more frequently
scrutinized by observers than the rugged and very interesting region which
includes these beautiful objects, selenographers can only give an incomplete and
unsatisfactory account of it...''

Paper 0 also contains a more quantitative treatment of the extent to which
observations of transients in Aristarchus might be significantly causally
correlated, suffice it here to say that there is little evidence of this,
before 1957.
This lack of signifcant correlation can also be considered an ``integral
constraint'' on the importance of observer preconception as to the existence of
TLPs as an important factor (for Aristarchus, at least) in determining the
observation selection function; furthermore, they provide no evidence for a
``hysteria signal'' of false reports due to special attention.
Elger's statement above implies that the ratio of observing time for
Aristarchus and the plateau versus equal areas not near the limb is at least of
order unity, and probably more.
We will see on the basis of $^{222}$Rn alpha particle measurements from Apollo
and {\it Lunar Prospector} in sections below that this cannot with any
reasonable probability imply that TLPs occur all over the Moon at the rate
reported as in Aristarchus (and hence we are not simply being fooled because
human observers spend more time looking at the Aristarchus plateau).

There is a pause in the frequency in TLP reports in both the Cameron (1978) and
Middlehurst (1968) catalogs, and indeed the break in reports 1927-1931 divides
the Middlehurst catalog at the median epoch in the catalog.
I will exploit this to compare both halves of the sample and eliminate 
over-reporting artifacts by rejecting the higher of the two counts for a given
lunar feature in the manner that one can use to remove artifacts from two
exposures in a sequence of the same picture with a poorly defined non-Poisson
noise component.
Specifically, I bin the counts seen in Figure 1 into 300 km square ``pixels''
and take the smaller of the two counts for each pixel from before and after
1930, producing Figure 2.
Since each pixel can be labeled with the name of the feature(s) identified by
the observers in the reports that filled that pixel, I list the corrected count
for each feature or group of features (Table 2).
Within each pixel, I re-evaluate particular features to see if TLPs from the
two samples truly correspond geographically.
If TLPs occur in the same named feature (and I include any positional
information available), or within a 50 km radius of each other, or within
1.5$\times$ the radius of the named crater, whichever is larger, I retain this
as a match.
The latter is a rejection consideration in less than 10\% of the cases.
This resulting count from this entire procedure is likely to be much more
robust against selection biases than the distribution shown in Figure 1, or for
that matter similar plots shown by previous authors who did not impose an
artifact rejection algorithm.
I am assuming in effect that there are quantitatively different observing
strategy results during these two time periods, which are capable of producing
spurious peaks in the geographic distribution of reports, but do not completely
neglect any area of the nearside Moon, excepting geometric effects such as
limb foreshortening or lunar phase selection due to evening/morning viewing
times, which are independent of time when averaged over the libration period
(between one day and one sidereal month).
My appraisal of the literature is that this is probably a good assumption.

In some cases reports are tied only to individual mare as features, which are
larger than a pixel.
The impact of this systematic uncertainty is small, only two cases with one
report apiece, which I do not plot in Figure 2.
These correspondences are probably spurious, and I do not
include them in our mare/highland boundary discussion below, although I include
them in Table 2.

\noindent
Table 2: Number of TLPs Reported Per Feature, Corrected for Possible Artifacts
\begin{verbatim}
__________________________________________________________________________
Robust
Report   Feature(s)
Count
-------  -----------------------------
66       Aristarchus/Vallis Schroteri
15       Plato      
 2       Grimaldi
 2       Messier
 1 each  Alphonsus, Bessel, Cassini, Copernicus, Gassendi, Kepler,
         Lichtenberg, Littrow, Mare Humorum, Mare Nubium, Mons Pico, Pallas,
         Picard, Ptolemaeus, Riccioli, South Pole, Theaetetus, Tycho
__________________________________________________________________________
\end{verbatim}

Note that the Aristarchus plateau persists as the prime TLP site with 63\% of
the corrected report count total (of 104), but Alphonsus and Gassendi have
virtually disappeared (with one), and Ross D and Mare Crisium are gone
altogether.
Alphonsus in particular involved reports (except one) only since the Alter
(1957) observations, which precipitated a great deal of amateur interest.
Beyond Aristarchus, Plato is still a prominent feature with 15 counts, but
besides these two craters only Grimaldi and Messier survive with more than one
count (having only two apiece).
If the frequency of TLPs at a given site varies radically on the timescale of
centuries down to a few decades, features might drop from Table 2.
This selection filter is meant to sacrifice completeness in this case for
reliability.
Depending on the long-term fluctuations in TLP behavior, there may be
additional, significant TLP sites than what appears in Table 2.
For the sake of further discussion in this paper I assume the rates are
constant on these timescales.

Plato is a distinct, flooded crater on the northwestern edge of Mare Imbrium,
so is about 3.5 Gy old or older.
It sits near mountainous regions such as Montes Alps, and appears very dark in
comparison, and is very different than Aristarchus in visual appearance.
It can be striking in its long shadows stretching across its face when near the
terminator.
Some observer descriptions sound suspiciously like reports of this normal
activity, but most do not correspond to normal appearance (see Haas 2003).
In 1854-1889 there were four reports involving at least some experienced 
observers noting extremely bright point sources that appeared for 30 min up to
5 h (the longest duration report considered here); it is unclear if these
reports might have influenced each other.
There are few reports involving red sources (3 not during eclipse); there are
many reports of cloud-like appearance.

In detail, if a feature is reported in an unbiased way, one should expect the
count $N_1$ in Table 1 related to $N_2$ in Table 2 by $N_1=2(N_2+\sqrt{N_2/3})$
on average, for the case of taking the lowest of two values deriving from the
same Poisson distribution.
For Aristarchus $+$ Vallis Schr\"oteri $+$ Cobra-Head $+$ Herodotus, the total
in Table 1 is $N_1 = 150$, whereas $2 (N_2 + \sqrt{N_2/3}) = 137.4$, so the
comparison is consistent with a fraction $0.916 \pm 0.078$ of reports being
real.
This is 86\% for Plato, and essentially 100\% for Grimaldi and Messier (within
the limits of small number statistics).

This implies that approximately 70 events should have been detected in the
Aristarchus plateau before 1930 at the intensity at which the Moon was
observed during that interval.
Since this represents approximately half of the TLP reports during this time,
during which most reports occurred between 1700 and 1930, it seems consistent
with approximately one TLP per two years across the sample.
The rate since 1930 for the Aristarchus region is about four times the report
rate prior to this, and it is unclear how much of this is real increase in
event coverage versus false detections.
It may be simply the effect of the production of many, inexpensive telescopes.
Taking the pre-1930 rate just inferred as a lower bound and adjusting for the
fact that the Moon is only observable about 20\% of the time from the places
where observers were posted (accounting for Sun/Earth/Moon position and
weather), it seems TLPs occur at least twice yearly on average, approximately.
The corresponding rate after 1930, which might have an observing duty cycle
closer to unity, but might still suffer from residual spurious reports, is
about once per month.

In Paper 0, I perform additional robustness tests largely independent of this
one, requiring consistency by (1) taking the median of four comparably-sized
historical subsamples (before year 1877, 1877-1930, 1930-1956, after 1956), or
by taking the median of just the first three subsamples, (2) taking the median
over the season of year of the TLP report, before 1956, and (3) the median over
subsamples grouping the reports by geographical location of the observer,
before 1956.
Despite that these should be different in their sensitivity to observer bias
and error, they nonetheless give similar results: Alphonsus, as well as Ross D
and Gassendi, largely disappear; Aristarchus remains as by far the strongest
signal, followed by Plato (about three times weaker).
To slight degree recent impacts Tycho, Kepler and Copernicus become stronger
signals in these other tests.
Even most of the weak features in Table 2 remain; Eratosthenes occasionally
appears at a slightly stronger level.
Mare Crisium is the only signal to vary significantly in strength between the
different robustness estimates, in some cases reaching half the strength of
Plato.
Since it is actually two ``pixels'' in diameter, I am unsure that this should
even be included as a feature in this analysis.
On the whole, however, the consistent behavior of the main features in the
sample lends credence to the notion that this approach has some validity.
We are testing whether given features are robust either in human observing
behavior, or in the long-term variability of the actual physical processes
producing TLPs at given sites.
At least I have varied the former in several significant ways and find its
effects to be consistent for most features, and inconsistent primarily in those
features where history casts some suspicion.

Figure 2 as well as Figure 1 appears to retain the property that the points are
clustered around the mare/highland interface.
To develop the locus for this boundary is a challenge, but guided by the
observation by Li and Mustard (2000) that the highlands and maria have distinct
compositions and that this is immediately apparent in UV/visible flux ratio
maps such as those available from {\it Clementine} (see also Whitaker 1972 for
UV/IR).
We would like to develop a statistical test exploiting the separation between a
given TLP site and the closest boundary segment.
This depends on not only the length of this curvilinear boundary but also its
Hausdorff index (as in a Mandelbrot set) and flux ratio threshold, somewhat
arbitrarily (see Appendix I).
I intend to explore this further, but for now a simple hand-drawn curve based
on {\it Clementine} maps indicates that the points in Table 2 (weighted
by report count) are about 7 times closer to the boundary than random points,
which is a statistically significant result (at the $\sim 99.999$\% level).
This TLP correlation still suffers from the objection that some observer effect
might manufacture reports at the mare/highland boundary, however, even after
circumvention of the fractal/threshold problem.
When I remove the points in Table 2 from Table 1 and correlate the residuals,
I get a $2.5\times$ greater closest boundary separation, but this is for more
points and hence significant at a very high level.
Whatever is causing the TLP/boundary correlation appears to survive even in the
points that did not pass the more robust TLP report filter, so there appears to
be a residual effect of this mechanism, whatever it is, in the rejected points.
A natural explanation might be that many of the less active points are real,
but create a TLP sufficiently rarely so as to not repeat over decades or even
centuries, in which case the total TLP rate might be doubled or more.

The maria/highland boundary signifies several additional geophysical and
mineralogical
transitions: the change in albedo and UV/IR and UV/optical properties already
mentioned -- which is tied to composition; an apparent correlation with rille
structures corresponding to lava flows draining into mare basins, presumably
(Whittaker 1972); and even changes in electrical conductivity properties
presumably related to deep basalt concentrations and differing structure and
cooling due to the ancient presence of lava (Vanyan et al.\ 1979).
The cooling of the maria and highlands were very different (Reindler \&
Arkani-Hamed 2001), which might lead to a situation in which mascons that tend
to underlie the maria that were supported at early times might come to strain
the surrounding material as the maria cool.
Since the highlands are heavily fractured, while the maria are more
``annealed,'' the mare/highland boundary might also be the location where
basalt-entrained gas might most easily escape.
Most importantly, as we shall see below, there is significant evidence for
enhanced outgassing at the mare/highland interface, and this, I speculate,
might be due to the release of trapped maria gas, treated in more
detail in \S 3.3.

\subsection{Controversy over The Reality of TLPs}

I deal in Paper 0 with several works considering explanations of TLPs as
non-lunar or non-physical (usually observer-effect) mechanisms.
To summarize here, none of these seem to explain more than a small minority of
TLP reports, although one or two issues are left as loose ends.

Any scientist should be skeptical of any conclusion based solely upon the
existing optical data base of TLP reports.
Most of them are anecdotal, not independently verified, and involve no
permanently recorded signal that did not pass through the human visual cortex.
Many of the observers are not professional, and some are not even very
experienced.
Our results above indicate that a significant number are of inconsistent rates,
and might be spurious.

The onus of the argument must burden those who would convince us that TLPs are
real.
When it comes to locating a spurious effect that might explain the bulk of TLP
reports as unrelated to the vicinity of the Moon, absence of evidence is not
evidence of absence.
Given the inability heretofore to test a reported TLP in a timely manner with
sufficiently complementary measurements, we must ask if any other physical
effects firmly tied to the lunar environment are correlated with TLPs.

A investigation by Cameron (1967, 1972) and Middlehurst (1977a, b) into
correlations with several possible lunar parameters turn up primarily null
relations e.g., lunar anomalistic period (time between perigees), and lunar age
(phase), and find some correlation with perigee and crossing of the Earth's
magnetopause and bow shock, plus a strong correlation with local sunrise which
might be a selection effect based on observers' attraction to this area of
higher contrast.
Middlehurst (1977a, b) also claims a statistically significant positional
correlation between TLPs and shallow moonquakes (from Nakamura et al.\ 1974),
which separately have been tied to $^{40}$Ar release (Hodges 1977, Binder
1980).

One transient phenomenon which occurs on a regular basis is the elevation of a
tenuous dust layer at the local shadow terminator as observed by Lunokhod-II
(Severny et al.\ 1975) and Surveyor 7 (Gault et al.\ 1968a, b, Rennilson 1968)
(and perhaps detected extending to high altitudes by astronauts on Apollos 10,
17 and perhaps 8 and 15 - Criswell \& Freeman 1975) , which Criswell (1972)
ties to electrostatic dust elevation at the terminator caused by photoelectric
ejection in daylit areas creating a voltage up to 550 V within about 1 cm of a
shadow's edge.
Few TLPs are consistent with this mechanism, however.
For the remainder we need to find some mechanism\footnote{See Hughes (1980) for
a short review.}
to create such a disturbance near the lunar surface if TLPs are to be believed.
Paper II will deal with the details of such candidate mechanisms.
There are other transient processes occurring on the Moon, and it is the
primary purpose of this Paper I to ask if there is any such tie-in to TLPs.

\section{Lunar Outgassing}

\subsection{Geological Evidence of Trapped Lunar Gas}

Lunar sample evidence, include basalt vesicles and volatile coatings, indicates
that the eruption of mare lavas came with the release of copious amounts of
gas, although the nature of such gas is still somewhat mysterious.
Mare basalts that were exposed to the surface are riddled with a large volume
filling-factor of voids or vesicles (for a review see O'Hara 2000, some
examples are {\it Apollo 15} sample 15556 and {\it Apollo 17} sample 71155).
The volatiles whose pressure produced these vesicles are unknown; some
candidates have been modeled based on lunar petrology and knowledge of
terrestrial basaltic volatile content: CO, COS, Na, SO$_2$, S$_2$, in
decreasing order of likely concentration (Sato 1976) and probably CO$_2$.
Wilson \& Head (2003) discuss possible concentration levels of various gases,
but with considerable uncertainty.
Unfortunately, measuring the amount of gas once trapped in the vesicles or
inferring its density and content is difficult (O'Hara 2000).
If volatiles were trapped in the basalt, they most likely escaped (although
even this is controversial, c.f.\ Taylor 1975).
Circumstantial evidence has been found recently for endogenous water in some
lunar minerals (McCubbin et al.\ 2007).

In lunar fines carbon/nitrogen compounds are found primarily as CO, but also
CO$_2$, CH$_4$, and traces of HCN, C$_2$H$_2$ and N$_2$, as well as trace
O$_2$, cumulatively at about 200 ppm (Burlingame et al.\ 1970, which did not
treat N compounds explicitly).
Most of this gas might be due to reactions of solar-wind implanted atoms
(Hodges et al.\ 1973b).

One must consider the actions of fire fountains driven by gas into the vacuum
(see Biggar et al.\ 1971, 1972).
Evidence for such fire fountains is found in the orange glass and crystallized
black beads in {\it Apollo 17} samples (Elkins-Tanton et al.\ 2005).
Inclusions in these beads offer one way of sampling the ancient volatile
content of the magma (Ebel et al.\ 2005).
One recent paper gives convincing evidence that highly volatile substances
were contained in the formation of fire-fountain glasses (Saal et al.\ 2007),
including H$_2$O, F, S and in most cases Cl (but not CO$_2$), with H$_2$O being
detected at levels of $\sim 4-50$ ppm ($\pm 1$ ppm).
The possible connection between former lunar activity and possible transients
observed now has not been ignored (Friesen 1975, Classen 1974).
The outgassing/TLP connection has not been established on the basis of the
petrological record, however.

\subsection{Apollo Mass and Ion Spectrometers}

The tentative but intriguing nature of our knowledge of lunar outgassing is
summarized by Srnka (1979), and its ambiguity is impressed by Freeman and
Benson (1977).
It is reasonably clear that $^{40}$Ar is released by moonquakes (Hodges 1977,
Binder 1980), not predominantly solar wind implantation (Hodges et al.\ 1974a).
Also, bursts of gas, from neither artificial nor extra-lunar sources, have been
recorded coming from near the lunar surface.
Hodges et al.\ (1973a, b, 1974b) report a burst recorded by the {\it Apollo 15}
orbital mass spectrometer (at UT 1971 August 6, 08:22), showing species of 14,
28 and 32 amu, N$_2$ and perhaps O$_2$, near the northwest edge of Mare
Orientale just on the far side (110$^\circ$.3W, 4$^\circ.1$S) -- hence would
not be seen from Earth if a TLP, but Hodges et al.\ rule out many anthropogenic
mechanisms.
This burst was so rapid that the scanning mass spectrometer was incapable of
covering all species, but it is estimated that at least 10 kg of gas was
involved.\footnote{
On the surface the {\it Apollo 17} mass spectrometer (Hodges et al.\ 1973c)
recorded a burst (at UT 1973 February 22 22:30) which included N$_2$, NH$_3$
and perhaps ethane.
The release is thought to contain 10-50 kg of gas and originate from a source
100-300 km from the {\it Apollo 17} landing site (Criswell \& Freeman 1975,
transmitting unpublished report by R.\ Hodges).
Hodges et al.\ (1973b) do not include this event in their sample, however.}
Freeman et al.\ (1973) report similar bursts of OH$^{^{-}}$ ions recorded
by the {\it Apollo 14} ALSEP Suprathermal Ion DEtector (SIDE).
Both the Freeman et al.\ (1973) and Hodges et al.\ (1973c) reports were
re-evaluated two decades later (Freeman \& Hills 1991, Hodges 1991), although
not in light of new data, and doubt cast on their non-artifical nature.

The ALSEP mass spectrometer at the {\it Apollo 17} site indicates
that radiogenic $^{40}$Ar is released episodically, which is puzzling unless
there is venting from deep within the Moon (Hodges \& Hoffman 1974).
Importantly, the ALSEP mass spectrometer provides evidence that the Moon
releases CH$_4$, and perhaps other molecules from its surface at a local
molecular number density of $\sim$6000~cm$^{-3}$ over a 25 hour period at
sunrise (Hoffman \& Hodges 1974, 1975).
Most of these signals are small, of marginal or slightly higher statistical
significance ($3\sigma$ for CH$_4$, $2\sigma$ for NH$_3$, $1-2 \sigma$ for
H$_2$O, CO, and CO$_2$: Hoffman \& Hodges 1975.
N$_2$ and O$_2$ as seen in the burst from orbit are at the $1\sigma$ level.)~
The presence of these molecules at all, even if at tiny concentrations, is
cause to suspect an outgassing source, since the sum of concentrations of H, N
and C in all forms in the regolith totals only about 200-300 ppm.

One point stressed in Papers II and III is that a simple model indicates
that the amount of outgassing needed to sustain this monthly volcanic signal is
of order 10-30 tonne y$^{-1}$, similar to the propellant load anticipated
for the Lunar Surface Access Module (17 tonnes).
If we are going to isolate and study sources of lunar outgassing, it will be
much easier before the bulk of human exploration on the Moon.

\subsection{Orbiting Alpha-Particle Spectrometer: {\it Apollo 15, 16 \& Lunar
Prospector}}

The crust of the Moon contains about 20 ppb of uranium (Drake 1986), mostly
$^{238}$U which decays eventually to $^{222}$Rn in $4.5 \times 10^9$y
(half-life).
Over the thickness of the lunar crust of 64 km (Zuber et al.\ 1995), this
implies that the Moon produces $\sim 10$ g s$^{-1}$ of $^{222}$Rn, assuming
these values pertain homogeneously, which corresponds to a decay rate density
of 40 cm$^{-2}$ s$^{-1}$ assuming all $^{222}$Rn reaches the surface.
How much of this escapes to the surface?
(Simple diffusion is not important: Friesen \& Adams 1976.
Also, see Hodges (1975) for an alternative analysis based in analogy to
$^{40}$Ar from $^{40}$K decay).

The way to establish this would be with orbiting alpha particle spectrometers
of the kind that were flown on {\it Apollo 15} (Gorenstein et al.\ 1974), {\it
Apollo 16} (Golub et al.\ 1973) and {\it Lunar Prospector} (Lawson et
al.\ 2005).
The global {\it Lunar Prospector} $^{222}$Rn decay map averages about 0.004
cm$^{-2}$ s$^{-1}$, which amounts to $2 \times 10^{15}$ s$^{-1}$ or $\sim
2$ g y$^{-1}$.
Most of the $^{222}$Rn produced in the crust does not leak out, but $10^{-4}$
of it does, within the half-life of 3.8 d.
This amounts to the equivalent of the outer 20 m or so of the regolith (roughly
its typical depth), which does not bespeak leakage from the deep crust,
seemingly.
(Either this, or the gas takes typical 50 d to reach the surface from anywhere
in the crust.)

In spite of the above calculation, the detailed structure of the $^{222}$Rn
decay map implies a more complicated situation which does seem to indicate
involvement with the deeper Moon.
In orbit these instruments will see alphas flying in nearly straight-line
paths from their decay site (deflected slightly by magnetic fields), with a
locational accuracy comparable to the elevation of the spacecraft (for one
alpha) but to better accuracy for a point source if it is strong enough to be
centroided using multiple detections.
The orbiting alpha particle spectrometer on {\it Apollo 15} and {\it 16}
revealed two types of features, against a nearly constant background level:
1) a consistent enhancement of the alpha particles of $^{^{210}}$Po, a
daughter product of $^{^{222}}$Rn gas, over the maria edges (Gorenstein et
al.\ 1974), and
2) anomalous outbursts of $^{^{222}}$Rn $\alpha$ particles (only 3.8~d
half-life) probably tied to recent outgassing, over craters Grimaldi and
Aristarchus (Gorenstein et al.\ 1973, Gorenstein \& Bjorkholm 1973), both
prime TLP sites.
Gorenstein et al.\ conclude that since $^{^{210}}$Po and $^{^{222}}$Rn were
not in radioactive equilibrium over these sites, radon must have been released
sporadically and recently, with large amounts within the past few decades.

{\it Lunar Prospector} orbited the Moon during 1998 January 17 to 1999
October 25, {\it Apollo 15} during 1971 July 29 to August 4, {\it Apollo 16}
during 1972 April 19-25, and all three stayed about within 93-120 km above the
surface.
{\it Lunar Prospector} covered the whole lunar surface, whereas {\it Apollo 15}
was confined to a strip within $28^\circ$ of the equator, and for {\it Apollo
16}, $12^\circ$ (which took {\it Apollo 15} over Aristarchus but not
{\it Apollo 16}; neither over Plato).
These detectors were sensitive to the decay of $^{222}$Rn (half-life of 3.8 d),
its daughter $^{218}$Po (3 m), half of which will be forced back into the
regolith from its recoil from $^{222}$Rn gas decay and half lost to space,
within the $^{222}$Rn migration radius of a few hundred km, and $^{210}$Po
(21 y, effectively due to the decay of $^{210}$Pb).
{\it Lunar Prospector} likewise detected a $^{^{210}}$Po mare edge correlation,
and two episodic $^{222}$Rn releases.

The {\it Apollo 15} and {\it Lunar Prospector} alpha-particle spectrometers
detected at least four signals from recent episodic activity: at Grimaldi,
Kepler and twice from Aristarchus (the {\it Apollo 15} and {\it Lunar
Prospector} events being sufficiently separate in time to be effectively
independent despite their positional coincidence).
Note that the {\it Lunar Prospector} signals were time-averaged over the
mission, so may indicate more than one event apiece (for Aristarchus and
Kepler).
It is notable that all are on the near side, location of nearly all maria.
If TLPs were exactly correlated with $^{222}$Rn, the results shown in Figure 2
and Table 2 would predict that given four uncorrelated events chosen at
random, the most probable result would be two events on the Aristarchus
plateau, and two events distributed among any of the features on the list
(favoring Plato except for the fact that it is too far north to have been seen
easily by {\it Apollo 15} and not in the range of {\it Apollo 16}).
This describes exactly what is observed.
A simple non-parametric test comparing the distribution of corrected TLP
counts versus alpha episodic activity, such as a two-sided Kolmogorov-Smirnov
test, provides a very low degree of rejection of the null hypothesis,
indicating that the sample distributions could easily be identical among these
features.
Furthermore, the fraction of the lunar surface represented by the sites listed
in Table 2 is very small, only about 11\% even if one includes each entire
``pixel,'' and much smaller if one restricts the area to the feature alone or
the region actually spanned by TLP reports for each feature.
Despite this, all four alpha episodes land within this area.
The episodic alpha-particle releases are extensive in area roughly on the
scale of a pixel, but they can be centroided better than this.
Given the state of the data set, I will not attempt to compute a realistic
correlation coefficient for the alpha versus TLP distributions, but it seems
very unlikely that these coincidences would occur at random, at roughly the
$10^{-4}$ probability level.
The orbit of {\it Lunar Prospector} was polar, while that of {\it Apollo 15}
was inclined $26^\circ$ to the equator (extending to $30^\circ$ in sensitivity
domain for alpha particles given the elevation of the spacecraft's orbit).
{\it Apollo 15} covers 67\% of the area of {\it Lunar Prospector}, but 73\% of
the TLP sites in Table 2, and 77\% of the TLP counts (54\% not counting
Aristarchus).
These fractional differences are not sufficient to change significantly
$P \la 10^{-4}$ for random radon/TLP coincidence.
This close correlation is even ignoring the result that within this $< 10$\%
of the Moon's surface covered by TLP-active ``pixels;'' the four $^{222}$Rn
events are distributed in a manner very similar to the TLPs.

Lower-level TLP activity seems to correlate with mare/highland edges, as does
the long-term signal for leakage of gas, for which $^{210}$Po represents a
proxy (see Appendix I).
The TLP/mare boundary correlation is very strong, while the $^{210}$Po signal
is limited by poor statistics to $P \la 10^{-4}$ of being non-random.
Nonetheless, this provides an independent statistical indication, separate from
the $^{220}$Rn result, that there is a correlation between TLP activity and
radon outgassing, even over long timescales.

If the TLP/$^{222}$Rn correlation is one-to-one, we can use the alpha particle
data to estimate TLP event rates (but we do not know how visible these
$^{222}$Rn events would be).
The Apollo and {\it Lunar Prospector} alpha particle spectrometers were in
orbit for a total of 293 days, compared to $\sim 10$ d in which an outgassing
event might remain within $10\times$ of full detectability.
During the $^{222}$Rn event time, the Apollo instruments would pass overhead at
least once, but Apollos 15 and 16 covered only about 45\% and 25\% of the
surface, respectively.
For {\it Lunar Prospector}, on a polar orbit, all points were covered, but
roughly 1/3 of the time.
This implies that an instrument covering the entire Moon 100\% of the time
might expect to detect about one every 24 d.
Assuming for the sake of simplicity that each of these produce a TLP, and TLPs
are only nearside phenomena (due to the farside paucity of maria), observers
should see about one per month at full observing duty cycle.
(This is just an approximation; one might speculate, for instance, that the
South Pole-Aitken might also produce TLPs.)

Since Aristarchus was the site of $^{222}$Rn episodes in both {\it Apollo 15}
and {\it Lunar Prospector} samples, the connection to TLPs has been noticed
already (Gorenstein \& Bjorkholm 1973, Lawson et al.\ 2005).
Uncertainty remained heretofore as to whether this might be due to an effect of
the extreme selection biases present in the TLP catalogs, but this doubt is
diminished for two reasons: 1) the TLP signal that I am discussing above
depends entirely on pre-{\it Apollo 15} TLP sightings, and the alpha
spectrometer surveys were highly unbiased, so there is no observer-based causal
link between the data sets;
and 2) the fact that a fair as possible treatment of the optical TLP selection
effects such as above causes the optical/$^{222}$Rn correlation to become even
more evident is a strong indication of their reality and association with
outgassing.
In addition to this a nearly equally strong correspondence between weaker TLP
sites and the long-term $^{210}$Po enhancement both tied to the mare/highland
boundary provides nearly independent and strong support to the tie-in of TLPs
and outgassing.

\section{Lunar Seismic Data and a Discussion}


At the outset, the ALSEP seismic record offers fascinating but confusing
insight into the physical nature of TLPs, in that it provides a third
dimension and enough information for considering physical mechanisms, and
appears to point to at least two.
I postpone most discussion of local physical mechanisms (such as changes in
albedo during explosive outgassing and coronal discharge) to Paper II, but
will touch on possibilities here for explaining the TLP spatial distribution.  

Why is gas leaking out of the Moon, preferentially at the maria edges and
around Aristarchus, a recent impact on an elevated region among maria?
There are several possibilities.
First and most simply, the mare/highland interface is the one place where the
fractured structure of the highlands interacts stratigraphically with more
structurally sound mare basalts.   
This leads to compositional boundaries and fractures in highlands materials
acting as channels to the surface for trapped gas related to mare emplacement.
In the case of Aristarchus, the pervasive volcanic conduits that fed the
materials that created the plateau act as channels for residual gases.
A second idea is that $^{40}$Ar might derive from high KREEP minerals since
buried by mare basalt, and that the quickest way for this gas to escape is via
subsurface migration in cracks below the maria, reaching the surface
concentrated at mare edges.
In this picture outgassing is potentially driven by purely radiogenic
production, not requiring recent volcanism.
This idea is troubled by the complex nature of igneous highland rock, which
presumably underlie the maria, that in some cases are high in KREEP
composition, but in many cases KREEP-poor (Simon \& Papike 1985).
It is uncertain which rock underlies the maria in question.
Also, this picture would presumably indicate large amounts of outgassing in the
highlands far from mare edges.

Hodges (1977) argues that $^{40}$Ar arises many hundreds of km deep below the
surface, with the outgassed $^{40}$Ar rate amounting to 3 tonne/y, about 6\%
of the total internal radiogenic production.
(This will be higher if a significant fraction is ionized, consistent with
SIDE results: Vondrak et al.\ 1974.)~
Runcorn (1974) proposes a model wherein episodic lava effusion can lead to the
production of mare mascons in layers denser than a single basalt mass, and that
cracks caused by the resulting strain of support can surround the maria and
extend through the lithosphere.
These can lead to moonquakes and also channels by which gas can escape perhaps
to produce TLPs (Friesen 1975, Runcorn 1977).
This is supported by the (weak) correlation of TLPs with maximal tidal stress
(Middlehurst 1977b).

To study this I look at the compilation of shallow moonquakes (Nakamura et
al.\ 1979) from the ALSEP seismograph array (concentrated on the equatorial
near side) and plot their locations (Figure 3).
With only 26 well-localized points, the
distribution at first appears random (excepting an overwhelming tendency for
events to congregate on the near side, with the greatest angular distance away
from Earth being only 11$^\circ$ onto the far side -- to some extent just a
sensitivity issue), visual inspection of Figure 3 indicates a tendency for
these to favor the maria and even their edges, as Nakamura et al.\ point out.
Again, I will calculate later a mare-edge correlational significance, which
depends on the threshold adopted for compositional differences maps which
effectively distinguish mare and highlands, and hence has a fractal nature that
needs further study in order to produce a result not affected by arbitrary
criteria, but for now I take at face value Nakamura et alia's statement, which
appears secure at the 99.9\% level based on the same mare/highland curve drawn
in \S 2.5.

Even though the entire sample of shallow moonquake loci is only 26 events over
eight years, one notes the total absence of the Aristarchus plateau from the
signal; either it was quiescent during this time, or gas leaks through its
cracks without being stimulated by strong moonquakes.
Given that this plateau contributes 61\% of reports in Figure 2, the spatial
distribution of shallow moonquakes differs from this at the level of $\sim 4
\times 10^{-8}$ probability of being a random result.
It may be that the massive impact which occurred at Aristarchus only $\sim$450
My ago has made the process of gas finding fractures to the surface easier; the
same might be said of Tycho, Copernicus and Kepler, the most prominent, recent,
nearside impacts\footnote{The age of Aristarchus is quoted variously as $450\pm
50$ My (Zisk et al.\ 1977) and $155\pm 25$ My (K\"onig et al.\ 1977), compared
to Copernicus: 900 My (Silver 1971), Tycho: 96 My (K\"onig et al.\ 1977,
Arvidson et al.\ 1976), and Kepler: $785\pm 160$ My (K\"onig et al.\ 1977) and
$75\pm 25$ My (Basilevsky et al.\ 1977).}, none of them on the mare/highland
interface but nonetheless prime TLP sites that survive the robustness sieve,
and with Kepler being a site of detected $^{222}$Rn outgassing.
This idea is perhaps borne out by the distribution of TLP report locations near
Aristarchus, which while not sampled uniformly, nonetheless seems to show
a concentration centered around the Aristarchus impact, rather than the whole
plateau itself.
(Of the 40 events near Aristarchus localized to within about 10 km, 11 are
contained within the 1500 km$^2$ of the crater itself, and all are within the
southwesternmost $10^4$ km$^2$ of the plateau, which totals $5\times 10^4$
km$^2$ in area.)~
There are no good shallow moonquake matches with any particular sites, beyond
the mare edge tendency.
(None of the total of 28 moonquakes, localized or not, land closer than 1.5 d
to a TLP report:\footnote{Cameron (1991) mentions the observation of the
temporal coincidence of a large moonquake and a surface water outgassing event,
but this appears to have not entered the refereed literature.}
on 1972 September 16/17 being the closest -- not statistically significant.)~
On the other hand, the rate of moonquakes is very similar to our estimate for
mare/highland boundary TLPs, which may not be totally a coincidence.

Somewhat paradoxically, deep moonquakes (Nakamura 2005), which are usually
thought to occur at depths (500 km$~\la depth \la 1500$ km) unassociated with
mare basalt plains e.g., Bulow, Johnson \& Bills (2006), are evidently
correlated with mare edges as well.
This is even a stronger result than for shallow quakes.
There are only two (or three) deep moonquakes near Aristarchus, out of a total
sample of 98; again the recent impacts Aristarchus, Tycho, Copernicus and
Kepler are not sites of major deep moonquake activity, while they are the sites
of TLP reports and $^{222}$Rn outgassing.
The correlation between the TLP reports shown in Figure 1 (or Figure 2) and
deepmoon quakes in Figure 3 is amazing, but there are limits to it: as well as
fresh impacts, the correlation around to Plato is diffuse at best, spread over
hundreds of km, and includes shallow events.

Moonquakes seem to be correlated with TLPs and presumably outgassing in terms
of the large-scale mare/highland boundary pattern, but not on a finer scale (in
space or time).
The two classes of events appear to be associated, but not directly correlated
in detail in a way indicating a prompt causal sequence.
A correlation does not guarantee a physical relation.
I will ask later whether this apparent smearing of the correlation might be due
to time delay or spatial dislocation, subsurface.
The presence of shallow moonquakes and outgassing events on the mare edges may
be a sign of the settling of mare basalt plains, as above.
This can be studied further by the examination of concentric fault or graben
structures (Lucchitta \& Watkins 1978), and is consistent with it.
In this case, typical mare plate edges are settling no more than about 100 m
over 3 Gy, or a rate under a few tenths of a $\mu$m per year around their
circumference.

Is this sufficient to release the observed gas?
As a simple model for illustration here, consider that the grinding front of
this mare slippage, if as long as the curve in Figure 2 ($\sim$10000 km), will
pulverize $\sim10^4 - 10^5$ tonne y$^{-1}$ of rock for each 100 km depth of
active fault, depending on the details of the slippage face.
From Wilson \& Head (2003), a reasonable estimate for the entrained gas
content might be $10^{-3}$ by mass.
This may liberate gas in large quantities; in Paper II, I will discuss
how much gas is needed to support the observational signatures discussed
above, tending toward 10-30 tonne y$^{-1}$, depending on how many and which
species.
The way in which this gas reaches the surface, how long it takes, and how much
it spreads from its source in the interior (as well as the total amount of gas
and what fraction thereof) will be regulated in part by the nature of this
grinding and how deeply it extends.

Of course we could also see gas leaking from elsewhere in the maria, not just
along the edges, if the settling (and impacts) cause them to fracture (which
they almost certainly do).
One final calculation is whether the mare-edge signal simply involves the
edge, or might involve fractures throughout, as one might suspect.
A glance at Appendix I would indicate that the later case probably dominates
for many of the datasets, although curiously perhaps not for deep moonquakes
and definitely not for TLPs that have passed the robustness test.
A larger dataset for both TLPs/outgassing and deep/shallow moonquakes will
help illucidate what mechanisms are in play.

What is the cause of TLP reports in major, fresh impacts Tycho, Kepler and
Copernicus?
Certainly they cause extended fractures, but their fractured/breccia lens
extends down only about 1/3 of the crater diameter (Hanna \& Phillips 2003),
and the fractures themselves less than the crater diameter (Ahrens et
al.\ 2002).\footnote{Diameters - Aristarchus: 42 km, Copernicus: 93 km, Kepler:
32km, Tycho: 85 km.}
These barely penetrate the crust, if at all, but do perforate the mare basalt.
Alternatively, Buratti et al.\ 2000 hypothesize that gas may be released by
avalanches down these young surfaces, or the outgassing itself may activate
mass wasting.

Aristarchus is unique in being about 30 times more active than any one of the
other three young craters; it is also the only such crater that arguably lands on
the mare/highlands boundary (the Aristarchus Plateau being highland-like both in
terms of both elevation and composition - although with differences in mafic
concentration: McEwen et al.\ 1994.)~
Regardless of this issue, the heightened activity level overlooks the singular
nature of the Plateau.
This region contains the largest density of sinuous rilles, and Vallis
Schr\"oteri is by many times the largest such rille on the Moon (Zisk et
al.\ 1977).
These are capable of having filled most of the volume of Oceanus Procellarum
(Whitford-Stark \& Head 1980).

The Procellarum KREEP terrane (PKT) is a unique region consisting primarily of
Oceanus Procellarum and Mare Imbrium (Jolliff et al.\ 1999, Haskin et
al.\ 2000), and may contain much of the thorium in the Moon, despite covering
only about 17\% of the surface.
While in the PKT maria Imbrium and Humorum correspond to large mascons (like
most of the large non-PKT maria: Serenitatis, Crisium and Nectaris),
Procellarum is notably mascon-free (see Konopliv et al.\ 2001).
Procellarum does not correspond to a well-defined, localized basins, and
following Runcorn (1974), is either shallow or was filled without solidifying
between lava effusion episodes.

It is worth noting that the South Pole-Aitken basin (SPA) is antipodal to the
eastern PKT.
The SPA center is quoted variously as 41$^\circ$.5S to 60$^\circ$S, and
174$^\circ$.5E to 180$^\circ$E (Hiesinger \& Head 2004, Leikin \& Sanovich
1985, Wood \& Gifford 1980), whereas the PKT centroid at 29$^\circ$N,
28$^\circ$W is 950 km away (depending somewhat on the background
level set for the Th $\gamma$-ray background: Lawrence et al.\ 1998); the SPA
and PKT are both roughly 2500 km across.
This recalls the hypothesis (Schultz \& Gault 1975) that the SPA impact should
produce a huge concentration of fractures on the opposite side of the Moon, in
the form of a radial column reaching from the deep interior to the surface (due
to spherical aberration effects in concentrating reflecting seismic waves).
Since the SPA is so large but shallow (12 km bottom to rim), and shows no
extreme compositional deviations from surface crust (Pieters et al.\ 1997,
c.f.\ Lucey et al.\ 1998), this leads to the hypothesis that the SPA impact was
a low-velocity, glancing blow (Schultz 1997), which can lead to an offset
between the initial impact (and resulting antipode fracture column) and the
actual center of the crater of many hundreds of km or even $\ga 1000$ km away.
There is no obvious disturbance at the apparent antipode (close to the southern
edge of Mare Frigoris near Plato), whereas Schultz (2007) would like to place
the antipodal fracture column near the center of the PKT to explain evidence of
recent outgassing (Schultz et al.\ 2001, 2006) and similar features.

We would argue that the center of the Aristarchus/Kreiger/Prinz volcanic
activity (at about 28$^\circ$N, 46$^\circ$W, 470 km from the PKT centroid, away
from the center of Imbrium), thought to be the major source of the PKT
effusion, is an even more likely SPA antipodal feature. 
Obviously, a large portion of the PKT originated within Imbrium crater, but if
the Terrane is due to SPA antipodal eruption, the Aristarchus activity (and to
a lesser extent Marius Hills) probably contributed much of the remainder.
One objection to both hypotheses, however, is the unknown direction of the
low-angle approach of the SPA impactor.
The thickest elevation of highlands material is to the north and east of SPA,
which would place the impact antipode to the north and west, although large
amounts of SPA ejecta must be found in the nearside southern highlands
(Peterson et al.\ 2002).
(If the SPA impact began opposite Aristarchus, ejecta would be thrown in the
direction of Mare Orientale - presumably a later impact which may have
redistributed much SPA ejecta.)~

If the SPA antipode eruption concept is valid, it seems likely to involve the
PKT and thereby the volcanic activity near Aristarchus.
If this is the case, the outgassing seen there now may be the residual of this,
connected with low-level activity plausibly involved with the deep Moon.

The further study of outgassing and the gas composition might offer many
insights into the lunar interior and evolution.
For instance, the fact that we see gas derived from heavier elements like 
uranium bespeaks only partial differentiation of the interior, which might be
probed additionally by understanding the behavior of very light volatiles, a
topic requiring much future work regarding TLPs.

\section{Summary and Conclusions}

In this paper, I study and cross-correlate various transient effects
occurring on the Moon: radon outgassing, moonquakes and optical transients.
The latter are somewhat problematic because they are the most heterogeneously
surveyed.
At the same time, this TLP database is much larger, offering the possibility that
we might remove the effects of observer selection bias and false reports.
This is worthwhile, because lunar outgassing, whether tied to TLPs or not,
would be a rare event, and the combined observational survey power of human
observers since the invention of the astronomical telescope would be by far
the most potent way to study these events if they are optically active.
While in the near future, robotic telescopes will supplant this database (see
Crotts et al.\ 2007), it is fortunate that I can produce consistent signals
from these data with a variety of robust sieves probing the structure of the
database in various ways.

The TLP data set is frought with selection effects and almost certainly at
least some false reports, for which explicit correction is problematic.
Nonetheless, the striking spatial correspondence between the distributions of
$^{222}$Rn episodic release and a sample of TLPs once they are culled of the
more obvious selection biases and bad data is strong evidence that lunar
outgassing is an important contributor to TLPs, with a probability at the
99.99\% level or greater.
Since there is little evidence in the TLP database and literature of a
``hysteria signal'' before 1956 which might be due to inexperienced or
overenthusiastic observers significantly polluting the sample with false
reports, the most likely systematic effect that might remain is overattention
to certain features by observers not seeking TLPs.
However, this cannot explain the geographical distribution of reports.
This is because the TLPs are confined to the same very small area as $^{222}$Rn
activity (hence they are almost certainly related), but TLPs are also highly
concentrated on Aristarchus (as may be $^{222}$Rn).
If the preponderance of Aristarchus reports were due to an observer selection
bias only, the implied amount of outgassing in the rest of the TLP region would
be at least two orders of magnitude greater than in Aristarchus alone (and more
than three orders of magnitude if extended to the entire near side).
As well as seeming increasingly implausible in terms of observer behavior, this
selection bias hypothesis would violate these physical constraints (the number
of $^{222}$Rn episodes detected being 4, not a few hundred or several thousand).
At least as it applies to Aristarchus, and presumably the rest of the sample,
much of the geographical structure must be due to real variation in TLP rates
near the lunar surface, not selection biases.

The related, but independent, correlation between lower-level TLP sites and
$^{210}$Po concentration is nearly as strong, and statistically (although not
physically) independent, indicating long-term as well as episodic correlation.
The $^{222}$Rn signal is almost certainly due to outgassing, because none of
the known effects associated with the mare/highlands interface listed in \S 2.5
would enhance $^{238}$U and therefore $^{222}$Rn (and therefore $^{210}$Po).
The radon must be transported to these regions, presumably mixed with other
gas, presumably through subsurface cracks.
The same applies to sites such as Aristarchus.
There may be other important mechanisms, but the evidence above indicates that
gas leaking from the Moon somehow changes the surface appearance in the optical
at least for limited periods of time.
These events appear to occur around Aristarchus and perhaps Plato, Grimaldi,
and recent impact craters, and may well occur at lower rates in a broad
distribution of locations.
TLPs can be used as a probe of lunar outgassing.

It appears that gas may leak out of the Moon for two reasons: because of the
tectonic sagging of the mare basalt, and some other mechanism that directs gas
out of impact fractures but does not produce detectable moonquakes.
Both may be in play at the Aristarchus Plateau and the latter at Kepler, Tycho
and Copernicus, all recent, major impacts there and elsewhere.
Surprisingly, there is an amazing correlation between the locus of TLPs not
including massive, fresh impact craters, and the distribution of deep moon
quakes.
The production of gas, and perhaps how this differs between these two kinds of
sites, has the potential of becoming a new way to dissect the lunar interior
structure and composition.

There may be a connection of TLP activity near Aristarchus to the massive
eruptions there that produced much of the Procellarum KREEP Terrane.
Since the PKT is antipodal to the South Pole-Aitken basin, the fractures
caused by the this giant impact on the other isde of the Moon may have provided
a channel for residual outgassing from the interior at Aristarchus.

In the following papers I will discuss the likely implications and possible
ways to enlarge our understanding of the connection between TLPs and lunar
outgassing.
In Paper II, I will discuss reasonable, simple models that help us understand
how gas might leak from the Moon and how that may produce TLPs.
In paper III, I propose several simple and powerful techniques which might be
exploited to learn about the internal structure, composition and evolution of
the Moon employing experiments involving observations from Earth and from the
vicinity of the Moon, and how these might relate to human activity there.

\section{Acknowledgements:} I would like to thank my colleagues Jani Radebaugh,
Denton Ebel, Caleb Scharf, Michael Joner, Daniel Austin, Patrick Cseresnjes and
Alex Bergier for useful discussion.
Also I thank Yosio Nakamura and the referees for their input.
I would especially like to express my appreciation to Winifred Cameron who
started me thinking about these issues.
Also, thank you to certain people who have listened to me reason through these
complex arguments and messy datasets.
(You know who you are!)

\newpage
\section{(Appendix I) - Calculations of Proximity to the Mare/Highlands
Interface}

As alluded to in the main text, the correlations of different samples with the
edges of the maria is an example of a Mandelbrot ``coast of Brittain'' problem
(Mandelbrot 1983), and is in particular sensitive to the smoothing scale, which
we will see below is a severe consideration in the case of the Aristarchus
plateau.
He have drawn a mare/highland boundary ``by hand'' aided by {\it Clementine}
albedo and UV/visible flux maps, as shown in Figure 2.
Compared to the locus several point distributions are correlated, and the
probability is calculated in two fashions as to the probability of this
correlation occurring at random:
1) measure the mean separation $d$ between a given point in the sample
distribution to the nearest segment of mare/highland boundary, divide this by
mean separation $m$ for a completely uniform distribution of points distributed
over the lunar surface, then raise this ratio $R = d/m$ (always less than one)
to the exponent equal to the number of points $n$, yielding a random
probability $P = R^n$.
This depends on the approximation that the points are close compared to the
size of the dominant structural scale in the boundary, hence is a
one-dimensional.  (The alternative, that the points are far away compared to
the size of boundary regions, has two-dimensional scaling, hence $P = R^{2n}$,
which is an even smaller probability).
This prescription is an approximation to a likelihood estimator where
$P = \Pi_{i=1}^n ~ d_i / m$,  where $d_i$ is the distance from the boundary for
each point.

Alternatively, 2) I must consider the change in $P$ if the maximum $d$ value
is removed (which measures the sensitivity to more such values), and consider
this as a $1 \sigma$ fluctuation in a Gaussian distribution.
This is usually the larger of the two probability
estimates for the chances of the result being random, and how much it would
change the mean  for the typical point to be removed from the distribution.
I cannot state this explicitly and concisely here since it depends on the
details of the distance distribution, but in all but two cases (F and H,
below), this is the larger of the two probabilities.

The several cases of mean closest separations versus the mare/boundary I
compute are: 

\noindent
A) Uniform Distribution over Near Side: mean separation $m_A = 7^\circ.9$ (as
measured in a great circle across the lunar surface), which I will use to
normalize most results below.

\noindent
B) Uniform Distribution over Both Sides: mean separation $m_B = 12^\circ.8$,
which reflects the much smaller number of maria on the far side.
This will be used in some cases to normalize the moonquake values.

\noindent
C) Uniform Distribution over Maria: mean separation $m_C = 5^\circ.4$, can be
used to establish if the correlation is with the edge of the maria versus the
entire mare area.

\noindent
D) Features in Table 1, Weighted ``Raw'' TLP Count, uncorrected by robustness
filter (and dominated by Aristarchus): $n=412$, $m_D = 1^\circ.5$, much smaller
than $m_A$, leading to a vanishing probability (1) above ($P_1 \approx
10^{-292}$), but a probability (2) corresponding to $25.7 \sigma$: $P_2 \approx
10^{-142}$, both ridiculously small and certainly overwhelmed by other effects
not treated here.

\noindent
E) Features in Table 2, Weighted by Robust TLP Count: $m_E = 1^\circ.1$, $P_2
\approx 7 \times 10^{-6}$, also depending heavily on whether the Aristarchus
plateau is counted as highland area.
(It has a partially consistent multispectral mineral signal.)

\noindent
F) Features in Table 2, Unweighted: $n=20$, $m_F = 5^\circ.5,$ $P_1 \approx 6
\times 10^{-4}$ is less sensitive to the Aristarchus plateau condition, but
effectively reduces $n$, so gives results only slightly weaker than (E).

\noindent
G) Features in Table 1 Unrepresented in Table 2, Weighted by ``Raw'' Count: $n
= 130$,
$m_G = 3^\circ.0$, $P_1 \approx 10^{-55}$ used to test if residual correlation
appears in the non-robust sample, which it obviously does, indicating some real
tendency of the remaining sample to cluster around the mare/highland interface.

\noindent
H) Shallow Moonquakes, Both Hemispheres: $n=26$, $m_H = 6^\circ.2$, $P_1
\approx 10^{-8}$, should be compared to $m_B$ and $m_C$, except for possible
shadowing at the ALSEP sites of some farside events due to a small molten core.

\noindent
I) Shallow Moonquakes, Near Side Only: $m_I = 5^\circ.3$,  $P_2 \approx 2
\times 10^{-4}$, but only three events need be dropped.

\noindent
J) Deep Moonquakes, Near Side Only: $n=98$, $m_J = 5^\circ.7$,  $P_2 \approx
10^{-14}$, which recovers the obvious visual impression that deep quake
loci follow the mare edges.
Note that the typical nearest-edge distance is comparable to the median
one-dimensional positional error of $4^\circ.7$ (avoiding the few anomalously
large values in the catalog), so the correlation may in reality be tighter.

I also study the distribution of $^{210}$Po from {\it Lunar Prospector}
(Lawson et al.~2005).
In their paper Lawson et al.~prefer to deal with statistically significant
potential sources ($2.2 \sigma$ to $3.8 \sigma$) rather than moments over the
entire $^{210}$Po distribution map, hence I will follow their preference in
ignoring low signal-to-noise pixels.
Note that there are 360 pixels total in this map, so $\sim$1.5\% of detections
are actually noise (less than one for $n = 13$, below).

\noindent
K) All $^{210}$Po sources: $n=13$, $m_K = 10^\circ.6$, $P_2 \approx 6.3 \times
10^{-5}$.
($P_1 \approx 0.088$ is a problematic overestimate given the size of spatial
bins in the $^{210}$Po map of Lawson et al.~(2005).
Correcting for this gives $m_K \approx 6^\circ .1$ and hence $P_1 \approx 6.5
\times 10^{-5}$.)

\noindent
L) $^{210}$Po sources, $>3\sigma$ detections: $n=6$, $m_J = 10^\circ.2$, $P_2 \approx 0.028$.

\newpage
\section{References}

\noindent
``Depth of Cracking beneath Impact Craters: Constraints for Impact Velocity''\\
\noindent
Ahrens, T.J., Xia, K.\ \& Coker, D.\ 2002, in {\it Shock-Compression of
Condensed Matter} eds.\ M.D.\ Furnish, N.N.\ Thadhani \& Y.\ Horie, (AIP: New
York), 1393.

\noindent
``A Suspected Partial Obscuration of the Floor of Alphonsus''\\
\noindent
Alter, D.\ 1957, PASP, 69, 158.

\noindent
``Making the photos of flashes on the Moon''\\
\noindent
Arkhipov, A.V.\ 1991, Zemlia i Vselennaya, 3, 76.

\noindent
``Cosmic Ray Exposure Ages of {\it Apollo 17} Samples and the Age of Tycho''\\
\noindent
Arvidson, R., Drozd, R., Guinness, E., Hohenberg, C., Morgan, C., Morrison,
R.\ \& Oberbeck, B.\ 1976, Proc.\ Lun.\ Sci.\ Conf., 7, 2817.




%
\noindent
``Photogeologic Study of Lunar Crater Rays: Nature of Rays and Age of Crater
Kepler''\\
\noindent
Basilevsky, A.T., Grebennik, N.N.\ \& Chernaya, I.M.\ 1977, Lun.\ Sci.\ Conf.,
8, 70.



%
\noindent
``Lunar lavas and achrondrites: petrogenesis of proto-hyperthene basalts in the
maria lava lakes''\\
\noindent
Biggar, G.M., O'Hara, M.H., Peckett, A.\ \& Humpries, D.J.\ 1971, in {\it 
Proc.\ Lunar Sci.\ Conf.}, 2, 617.

\noindent
``Maria lavas, mascons, layered complexes, achrondrites and the lunar mantle''\\
\noindent
Biggar, G.M., O'Hara, M.H., Humpries, D.J.\ \& Peckett, A.\ 1972, in {\it The
Moon}, eds.\ H.C.\ Urey \& S.K.\ Runcorn (IAU), p.\ 129.

\noindent
``Shallow moonquakes - Argon release mechanism''\\
\noindent
Binder, A.B.\ 1980, Geophys.\ Res.\ Let., 7, 1011.


%
\noindent
``The Lunar Craters Aristarchus and Herodotus''\\
\noindent
Birt, W.R.\ 1870, Astronomical Register, 8, 271.


%
\noindent
``Tidal Stress and Deep Moonquakes''\\
\noindent
Bulow, R.C., Johnson, C.L.\ \& Bills, B.G.\ 2006, Lunar Plan.\ Sci.\ Conf., 37,
abstract 1183.

\noindent
``Lunar Transient Phenomena: What Do the $Clementine$ Images Reveal?''\\
\noindent
Buratti, B.J., McConnochie, T.H., Calkins, S.B., Hillier, J.K.\ \&
Herkenhoff, K.E.\ 2000, Icarus, 146, 98.



%
\noindent
``Study of carbon compounds in Apollo 11 lunar samples''\\
Burlingame, A.L., Calvin, M., Han, J., Henderson, W., Reed, W.\ \& Simoneit,
B.R.\ 1970,
Geochim.\ et Cosmoch.\ Acta Sup., Volume 1. ``Proc.\ Apollo 11 Lunar
Sci.\ Conf., Vol.\ 2: Chemical and Isotope Analyses,'' ed.\  A.A.\ Levinson,
(Pergammon: New York), p.1779



%
\noindent
``Observations of changes on the Moon'' \\
\noindent
Cameron, W.S.\ 1967, 
Proc.\  5th Ann.\ Meet.\ Working Group on Extrater.\ Res., p. 47.

\noindent
``Comparative analyses of observations of lunar transient phenomena''\\
\noindent
Cameron, W.S.\ 1972, Icarus, 16, 339.


%
\noindent
``Lunar transient phenomena''\\
\noindent
Cameron, W.S.\ 1991, Sky \& Tel., 81, 265.



%
\noindent
``Degassing of the moon. III - The maria and interior of the moon''\\
\noindent
Classen, J.\ 1974, Solar System Research, vol. 8, no. 2, p. 70-73.

\noindent
``Lunar dust motion''\\
\noindent
Criswell, D.R.\ 1972, Proc.\ Lunar Sci.\ Conf., 2, 2671.

\noindent
``Interactions of the Interplanetary Plasma with the Modern Ancient Moon:
Summary of Conference''
Criswell, D.R.\ \& Freeman, J.W., Jr.\ \& 1975, The Moon, 14, 3.

\noindent
``Transient Lunar Phenomena: Regularity and Reality''\\
\noindent
Crotts, A.P.S.\ 2007,  ApJ, submitted.  (PAPER 0)

\noindent
``Lunar Outgassing, Transient Phenomena and The Return to The Moon, III:
Observational and Experimental Techniques''\\
\noindent
Crotts, A.P.S.\ 2007,  ApJ, submitted.  (PAPER III)

\noindent
``An Automated Lunar Transient Imaging Monitor on Cerro Tololo''
\noindent
Crotts, A.P.S., Hickson, P., Hummels, C.\ \& Pfrommer, T.\ 2007 (in
preparation).

\noindent
``Lunar Outgassing, Transient Phenomena and The Return to The Moon, II:
Predictions of Interactions between Outgassing and Regolith''\\
\noindent
Crotts, A.P.S.\ \& Hummels, C.\ 2007,  ApJ, submitted. (PAPER II)


%
\noindent
``Is lunar bulk material similar to earth's mantle?''\\
\noindent
Drake, M.J.\ 1986, in {\it Origin of the Moon}, eds.\ W.K.\ Hartmann,
R.J.\ Phillips, and G.J.\ Taylor, (Lunar \& Planet.\ Inst.: Tucson), p.\ 105.




%
\noindent
``Tomographic location of potential melt-bearing phenocrysts in lunar glass
spherules''
\\
\noindent
Ebel, D.S., Fogel, R.A.\ \& Rivers, M.L.\ 2005, Lunar Plan.\ Sci.\, 26,
No.\ 1505.

\noindent
``Selenographic Notes, February, 1884: Aristarchus amd Herodotus''\\
\noindent
Elger, T.G.\ 1884, Astron.\ Register, 22, 39.


%
\noindent
``Magmatic processes that produced lunar fire fountains''\\
\noindent
Elkins-Tanton, L.T., Chatterjee, N.\ \& Grove, T.L.\ 2003,
Geophys.\ Res.\ Let., 30, 1513.

\noindent
``Volatile Emission of the Moon; Possible Sources and Release Mechanisms''\\
\noindent
Friesen, L.J.\ 1975, Moon, 13, 425

\noindent
``Low pressure radon diffusion - A laboratory study and its implications for
lunar venting''\\
\noindent
Friesen, L.J.\ \& Adams, J.A.S.\ 1976, Geochim.\ et Cosmochim.\ Acta, 40, 375.

\noindent
``A search for gaseous emissions from the moon''\\
\noindent
Freeman, J.W.\ \& Benson, J.L.\ 1977, Phys.\ Earth \& Planet.\ Interiors, 14,
276.

\noindent
``The Apollo lunar surface water vapor event revisited''\\
\noindent
Freeman, J.W.\ \& Hills, H.K.\ 1991, Geo.\ Phys.\ Let., 18, 2109.

\noindent
``Observation of Water Vapor Ions at the Lunar Surface'' \\
\noindent
Freeman, J.W., Hills, H.K.\ \& Vondrak, R.R.\ 1973, Proc.\ Lunar Sci.\ Conf.,
3, 2217.

\noindent
``Post-sunset horizon `afterglow'"\\
\noindent
Gault, D.E.\ et al.\ 1968a, in {\it Surveyor 7 Mission Rep., 2}, JPL
Tech.\ Rep.\ 32-1264, p.~308.

\noindent
``Post-sunset horizon glow''\\
\noindent
Gault, D.E.\ et al.\ 1968b, in {\it Surveyor Final Science Rep., 2}, JPL
Tech.\ Rep.\ 32-1265, p.~401.


%
\noindent
``Distribution of $^{210}$Po Across the Apollo 16 Groundtrack and Correlations
with Lunar Surface Features''
\noindent
Golub, L., Bjorkholm, P.J.\ \& Gorenstein, P.\ 1973, Abs.\ Lunar \&
Plan.\ Sci.\ Conf., 4, 302.

\noindent
``Detection of Radon Emanation from the Crater Aristarchus by the Apollo 15
Alpha Particle Spectrometer'' \\
\noindent
Gorenstein, P.\ \& Bjorkholm, P.J.\ 1973, Science, 179, 792.

\noindent
``Detection of Radon Emanation from the Crater Aristarchus by the Apollo 15
Alpha Particle Spectrometer'' \\
\noindent
Gorenstein, P.\ \& Bjorkholm, P.J.\ 1973, Science, 179, 677.

\noindent
``Detection of Radon Emanation at the Edges of Lunar Maria by the Apollo Alpha
Particle Spectrometer'' \\
\noindent
Gorenstein, P., Golub, L.\ \& Bjorkholm, P.J.\ 1974, Science, 183, 411.


%
\noindent
``A New Model of the Hydrologic Properties of the Martian Crust and
Implications for the Formation of Valley Networks and Outflow Channels''\\
\noindent
Hanna, J.C.\ \& Phillips, R.J.\ 2003, Lun.\ Plan.\ Sci.\ Conf, 34, 2027.


%
\noindent
``The materials of the lunar Procellarum KREEP Terrane: A synthesis of data
from geomorphological mapping, remote sensing and sample analysis''\\
\noindent
Haskin, L.A., Gillis, J.J., Korotev, R.L.\ \& Jolliff, B.L.\ 2000, JGR, 105,
20.


%
\noindent
``Those unnumbered reports of lunar changes - Were they all blunders?''\\
\noindent
Haas, W.H.\ 2003, J.\ ALPO, 45, 2, 25.

\noindent
``Lunar South Pole-Aitken Impacr Basin: Topography and Mineralogy''\\
\noindent
Hiesinger, H.\ \& Head, J.W.\ 2004, Lun.\ Plan.\ Sci.\ Conf., 35, 1164.

\noindent
``Formation of the Lunar Atmosphere''\\
\noindent
Hodges, R.R., Jr.\ 1975, The Moon, 14, 139.

\noindent
``Release of Radiogenic Gases from the Moon''\\
\noindent
Hodges, R.R., Jr.\ 1977, Phys.\ Earth Planet.\ Interiors, 14, 282.

\noindent
Hodges, R.R., Jr.\ 1991, {\it personal communication}, in Stern, A.\ 1999,
Rev.\ Geophys., 37, 4.

\noindent
``Episodic release of $^{40}$Ar from the interior of the moon'' \\
\noindent
Hodges, R.R., Jr.\ \& Hoffman, J.H.\ 1974, 5th Lunar Sci.\ Conf., 3,
2955

\noindent
``The Lunar Atmosphere''\\
\noindent
Hodges, R.R., Jr., Hoffman, J.H.\ \& Johnson, F.S.\ 1974a,
Abstr.\ Lunar Planet.\ Sci.\ Conf., 5, 343.

\noindent
``The lunar atmosphere''\\
\noindent
Hodges, R.R., Jr., Hoffman, J.H.\ \& Johnson, F.S.\ 1974b, Icarus, 21, 415.

\noindent
``Composition and Dynamics of Lunar Atmosphere''\\
\noindent
Hodges, R.R., Jr., Hoffman, J.H., Johnson, F.S.\ \& Evans, D.E.\ 1973a,
Abstr.\ Lunar Planet.\ Sci.\ Conf., 4, 374.

\noindent
``Composition and Dynamics of Lunar Atmosphere''\\
\noindent
Hodges, R.R., Jr., Hoffman, J.H., Johnson, F.S.\ \& Evans, D.E.\ 1973b,
Proc. Lunar Sci.\ Conf., 4, 2855.

\noindent
``Lunar atmospheric composition results from Apollo 17''\\
\noindent
Hodges, R.R., Jr., Hoffman, J.H., Johnson, F.S.\ \& Evans, D.E.\ 1973c,
Proc. Lunar Sci.\ Conf., 4, 2865.

\noindent
``Methane and Ammonia in the Lunar Atmosphere'' \\
\noindent
Hoffman, J.H.\ \& Hodges, R.R., Jr.\ 1974, in ``Interactions of the
Interplanetary Plasma with the Modern and Ancient Moon,'' eds.\ D.\ Criswell \&
J.\ Freeman (Lunar Sci.\ Inst.: Houston), p.\ 81).

\noindent
``Molecular gas species in the lunar atmosphere'' \\
\noindent
Hoffman, J.H.\ \& Hodges, R.R., Jr.\ 1975, The Moon, 14, 159.

\noindent
``Transient Lunar Phenomena'' \\
\noindent
Hughes, D.W.\ 1980, Nature, 285, 438.





%
\noindent
``Thorium Enrichment within the Procellarum KREEP Terrane: The Record in
Surface Deposits and Significance for Thermal Evolution''\\
\noindent
Jolliff, B.L., Gillis, J.J.\ \& Haskin, L.A.\ 1999, in {\it Workshop on New
Views of the Moon}, eds.\ L. Gaddis \& C.K.\ Shearer (LPI: Houston), p.\ 31.



%
\noindent
``Photographic evidence of a short duration - Strong flash from the surface of
the moon''\\
\noindent
Kolovos, G., Seiradakis, J. H., Varvoglis, H.\ \& Avgoloupis, S.\ 1988, Icarus,
76, 525.

\noindent
``The origin of the Moon flash of May 23, 1985''\\
\noindent
Kolovos, G., Seiradakis, J.H., Varvoglis, H.\ \& Avgoloupis, S.\ 1992,
Icarus, 97, 142.

\noindent
``Recent Lunar Cratering: Absolute Ages of Kepler, Aristarchus, Tycho''\\
\noindent
K\"onig, B., Neukum, G.\ \& Fechtig, H.\ 1977, Lun.\ Plan. Sci.\ Conf., 8, 555.

\noindent
``Recent Gravity Models as a Result of the Lunar Prospector''\\
\noindent
Konopliv, A.S., Asmar, S.W., Carranza, E., Sjogren, W.L.\ \& Yuan, D.N.\ 2001,
Icarus, 150, 1.


%
\noindent
``Global Elemental Maps of the Moon: The Lunar Prospector Gamma-Ray
Spectrometer''\\
\noindent
Lawrence, D.J., Feldman, W.C., Barraclough, B.L., Binder, A.B., Elphic, R.C.,
Maurice, S.\ \& Thomsen, D.R.\ 1998, Science, 281, 1484.


%
\noindent
``Recent outgassing from the lunar surface: The Lunar Prospector Alpha Particle
Spectrometer''\\
\noindent
Lawson, S.L., Feldman, W.C., Lawrence, D.J., Moore, K.R., Elphic, R.C., Belian,
R.D.\ \& Maurice, S.\ 2005, JGR, 110, E09009.

\noindent
``Origin of the Southern Basin on the far side of the moon''\\
\noindent
Leikin, G.A.\ \& Sanovich, A.N.\ 1985, Astron.\ Vest., 19, 113.


%
\noindent
``Compositional gradients across mare-highland contacts: Importance and
geological implications of lateral transport''
\noindent
Li, L.\ \& Mustard, J.F.\ 2000, JGR, 105, 20431



%
\noindent
``Large Grabens and Lunar Tectonism''\\
\noindent 
Lucchitta, B.K.\ \& Watkins, J.A.\ 1978, Lun.\ \& Plan.\ Sci., 9, 666.

\noindent
``FeO and TiO2 concentrations in the South Pole-Aitken basin: Implications for
mantle composition and basin formation''\\
\noindent
Lucey, P.G., Taylor, G.J., Hawke, B.R.\ \& Spudis, P.D.


%
\noindent
``The Fractal Geometry of Nature''\\
\noindent
Mandelbrot, B.B.\ 1983, (Freeman: New York).

\noindent
``The Chinese Reader's Manual''\\
\noindent
Mayers, W.F.\ 1874 (Shanghai), p.\ 219

\noindent
``Is There Evidence for Water in Lunar Magmatic Minerals? A Crystal Chemical
Investigation''\\
\noindent
McCubbin, F.M., Nekvasil, H.\ \& Lindsley, D.H.\ 2007,
Lun.\ Plan.\ Sci.\ Conf., 38, 1354.

\noindent
``Clementine Observations of the Aristarchus Region of the Moon''\\
\noindent
McEwen, A.S., Robinson, M.S., Eliason, E.M., Lucey, P.G., Duxbury, T.C.\ \&
Spudis, P.D.\ \& 1994, Science, 266, 1858.



%
\noindent
``A survey of lunar transient phenomena''\\
\noindent
Middlehurst, B.\ 1977a, Phys.\ Earth Planet.\ Inter., 14, 185.

\noindent
``Transient lunar phenomena, deep moonquakes, and high-frequency teleseismic
events: possible connections''\\
\noindent
Middlehurst, B.\ 1977b, Phil.\ Trans.\ R.\ Soc.\ A.\, 285, 485.

\noindent
``Chronological Catalog of Reported Lunar Events''\\
\noindent
Middlehurst, B.M.,  Burley,, J.M., Moore, P.\ \& Welther, B.L.\ 1968, NASA
Tech.\ Rep.\ TR R-277.

\noindent
``Lunar transient phenomena: Topographical distribution''\\
\noindent
Middlehurst, B.M.\ \& Moore, P.A.\ 1967, Science, 155, 449.

\noindent
``The impact hazard''\\
\noindent
Morrison, D., Chapman, C.R.\ \& Slovic, P.\ 1994, in {\it Hazards Due to Comets
and Asteroids}, ed.\ T.\ Gehrels, (U.\ Arizona Press, Tucson).

\noindent
``Farside deep moonquakes and deep interior of the Moon''\\
\noindent
Nakamura, Y.\ 2005, J.\ Geophys.\ Res., 110, E01001. 

\noindent
``High-frequency lunar seismic events''\\
\noindent
Nakamura, Y., Dorman, J., Duennebier, F., Ewing, M., Lammlein, D.\ \& Latham,
G.\ 1974, Proc.\ Lunar Planet.\ Sci.\ Conf, 9, 3589.

\noindent
``High-frequency lunar seismic events''\\
\noindent
Nakamura, Y., Latham, G., Dorman, J., Ibrahim, A.-B.K., Koyama, J.\ \& Horvath,
P.\ 1979, Proc.\ Lunar Planet.\ Sci.\ Conf, 10, 2299.

\noindent
``Passive Seismic Experiment Long-Period Event Catalog''\\
\noindent
Nakamura, Y.\ et al.\ 1981, Galveston Geophys.\ Lab.\ Contrib., 491,
Tech.\ Rep.\ 18.



%
\noindent
``Flood Basalts, Basalt Floods or Topless Bushvelds? Lunar Petrogenesis
Revisited''\\
\noindent
O'Hara, M.J., 2000, J.\ Petrology, 41, 11, 1545.


%
\noindent
``Compositional Units on the Moon: The Role of South Pole-Aitken Basin''\\
\noindent
Peterson, C.A., Hawke, B.R., Blewett, D.T., Bussey, D.B.J., Lucey, P.G.,
Taylor, G.J.\ \& Spudis, P.D.\ 2002, Lun.\ Plan.\ Sci.\ Conf., 33, 1601.

\noindent
``Are there at present Active Volcanos upon the Moon?''\\
\noindent
Pickering, W.H.\ 1892, The Observatory, 15, 250.

\noindent
``The Moon: A Summary of the Existing Knowledge of Our Satellite, With a
Complete Photographic Atlas''
\noindent
Pickering, W.H.\ 1904, (Doubleday: New York).

\noindent
``Mineralogy of the mafic anomaly in the South Pole-Aitken Basin: Implications
for excavation of the lunar mantle''\\
\noindent
Pieters, C.M., Tompkins, S., Head, J.W.\ \& Hess, P.C.\ 1997,
Geophys.\ Res.\ Let., 24, 1903.


%
\noindent
``The Strength of the Lunar Interior''\\
\noindent
Reindler, L.\ \& Arkani-Hamed, J.\ 2004, Am.\ Geophys.\ Union, Fall 2001,
\#P12D-0524.

\noindent
``Sunset observations''\\
\noindent
Rennilson, J.J.\ 1968, in {\it Surveyor Final Science Rep., 2}, JPL
Tech.\ Rep.\ 32-1265, p.~119.

\noindent
``Some Aspects of the Physics of the Moon''
\noindent
Runcorn, S.K. \ 1974, Proc.\ Roy.\ Soc.\ London A, 336, 11.

\noindent
``Physical processes involved in recent activity within the moon''\\
\noindent
Runcorn, S.K. \ 1977, Phys.\ Earth \& Plan.\ Interiors, 14, 330.

\noindent
``The Volatile Contents (CO$_2$, H$_2$O, F, S, Cl) of the Lunar Picritic
Glasses''\\
\noindent
Saal, A.E., Hauri, H.\ \& Cooper, R.F.\ 2007, Lun.\ Plan.\ Sci.\ Conf., 28,
2148.

\noindent
``Indication of Luminescence Found in the December 1964 Lunar Eclipse''\\
\noindent
Sanduleak, N.\ \& Stock, J.\ 1965, PASP, 77, 237

\noindent
``Oxygen fugacity and other thermochemical parameters of Apollo 17 high-Ti
basalts and their implications on the reduction mechanism''\\
\noindent
Sato, M.\ 1976, Proc.\ Lun.\ Sci.\ Conf., 7, 1323.

\noindent
``Forming the South Pole-Aitken basin: The extreme games''\\
\noindent
Schultz, P.H.\ 1997, Lun.\ Plan.\ Sci.\ Conf., 28, 1259.

\noindent
``A Possible Link Between Procellarum and the South-Pole-Aitken Basin''\\
\noindent
Schultz, P.H.\ 2007, Lun.\ Plan.\ Sci.\ Conf., 38, 1839.

\noindent
``Seismic effects from major basin formations on the moon and mercury''\\
\noindent
Schultz, P.H.\ \& Gault, D.E.\ 1975, Earth, Moon \& Plan., 12, 159.

\noindent
``Recent Lunar Activity: Evidence and Implications''\\
\noindent
Schultz, P.H., Staid, M.\ \& Pieters, C.M.\ 2001, Lun.\ Plan.\ Sci., 31, 1919.

\noindent
``Lunar activity form recent gas release''\\
\noindent
Schultz, P.H., Staid, M.\ \& Pieters, C.M.\ 2006, Nature, 444, 184.

\noindent
``The measurements of sky brightness on Lunokhod-2''\\
\noindent
Severny, A.B., Terez, E.I.\ \& Zvereva, A.M.\ 1975, The Moon, 14, 123.


%
\noindent
``V-Th-Pb Isotropic Systems in {\it Apollo 11} and {\it Apollo 12} Regolithic
Materials and a Possible Age for the Copernican Event''\\
\noindent
Silver, L.T.\ 1971, Trans.\ Am.\ Geophys.\ Union, 53, 534.

\noindent
``Petrology of the {\it Apollo 12} Highland Component''
\noindent
Simon, S.B.\ \& Papike, J.J.\ 1985, JGR Suppl., 90, D47.

\noindent
``Lunar luminescence in the near ultraviolet''\\
\noindent
Spinrad, H.\ 1964, Icarus, 3, 500

\noindent
``Interior evolution of the Moon and the terrestrial planets''\\
\noindent
Spohn, T.\ 2004, in {35th COSPAR Ass'y, (Paris, France)}, p.~225.

\noindent
``On the detection of lunar volatile emissions''\\
\noindent
Srnka, L.J.\ 1979, Nature, 278, 152.

\noindent
Ina, a Lunar Caldera?
\noindent
Strain, P.L.\ \& El-Baz, F.\ 1980, Lun.\ Plan.\ Sci.\ Conf., 11, 1103.






%
\noindent
``Lunar Science: a Post-Apollo View''\\
\noindent
Taylor, S.R.\ 1975, (Pergamon: NY), 372 pp.

\noindent
``Electrical Conductivity Anomaly beneath Mare Serenitatis detected by Lunokhod
2 and Apollo 16 Magnetometers''\\
Vanyan, L.\ et al.\ 1979, Moon \& Planets, 21, 185.


%
\noindent
``The 1999 Quadrantids and the lunar Na atmosphere''\\
\noindent
Verani, S., Barbieri, C., Benn, C. R., Cremonese, G.\ \& Mendillo, M.\ 2001,
MNRAS, 327, 244.

\noindent
``Measurement of lunar atmospheric loss rate''\\
\noindent
Vondrak, R.R., Freeman, J.W.\ \& Lindeman, R.A.\ 1974,
Lun.\ Plan.\ Sci.\ Conf., 5, 2945.

\noindent
``Lunar Color Boundaries and Their Relationship to Topographic Features: A
Preliminary Survey''\\
Whitaker, E.A.\ 1972, in {\it Conf. on Lunar Geophys.}, (Lunar Sci.\ Inst.:
Houston), p.\ 348.

\noindent
``The Procellarum volcanic complexes: Contrasting styles of volcanism''\\
\noindent
Whitford-Stark, J.L.\ \& Head, J.W., III 1977, Proc.\ Lun.\ Sci.\ Conf., 8,
2705.

\noindent
``The Lunar Craters Aristarchus and Herodotus''\\
\noindent
Whitley, H.M.\ 1870, Astronomical Register, 93, 194.

\noindent
``Deep generation of magmatic gas on the Moon and implications for pyroclastic
eruptions''\\
\noindent
Wilson, L.\ \& Head, J.W.\ 2003, Geophys.\ Res.\ Let., 30, 1605. 


%
\noindent
``The Moon in Ultra-Violet Light, and Spectro-Selenography''\\
\noindent
Wood, R.W.\ 1910, MNRAS, 70, 226

\noindent
``Evidence for the Lunar Big Backside Basin''
\noindent
Wood, C.A.\ \& Gifford, A.W.\ 1980, in ``Conf.\ on Multi-ring Basins
(Abstracts)'' (LPI: Houston), p.\ 121.

\noindent
``The Aristarchus-Harbinger Region of the Moon: Surface Geology and History
from Recent Remote-sensing Observations''\\
\noindent
Zisk, S.H., Hodges, C.A., Moore, H.J., Shorthill, R.W., Thompson, T.W.,
Whitaker, E.A.\ \& Wilhelms, D.E.\ 1977, Moon, 17, 59.

\noindent
``Gravity, Topography and the Geophysics of the Moon from the Clementine
Mission''\\
\noindent
Zuber, M.T., Smith, D.E., Lemoine, F.G.\ \& Neumann, G.A.\ 1995, Science, 266,
1839. 

\newpage


\begin{figure}
\plotfiddle{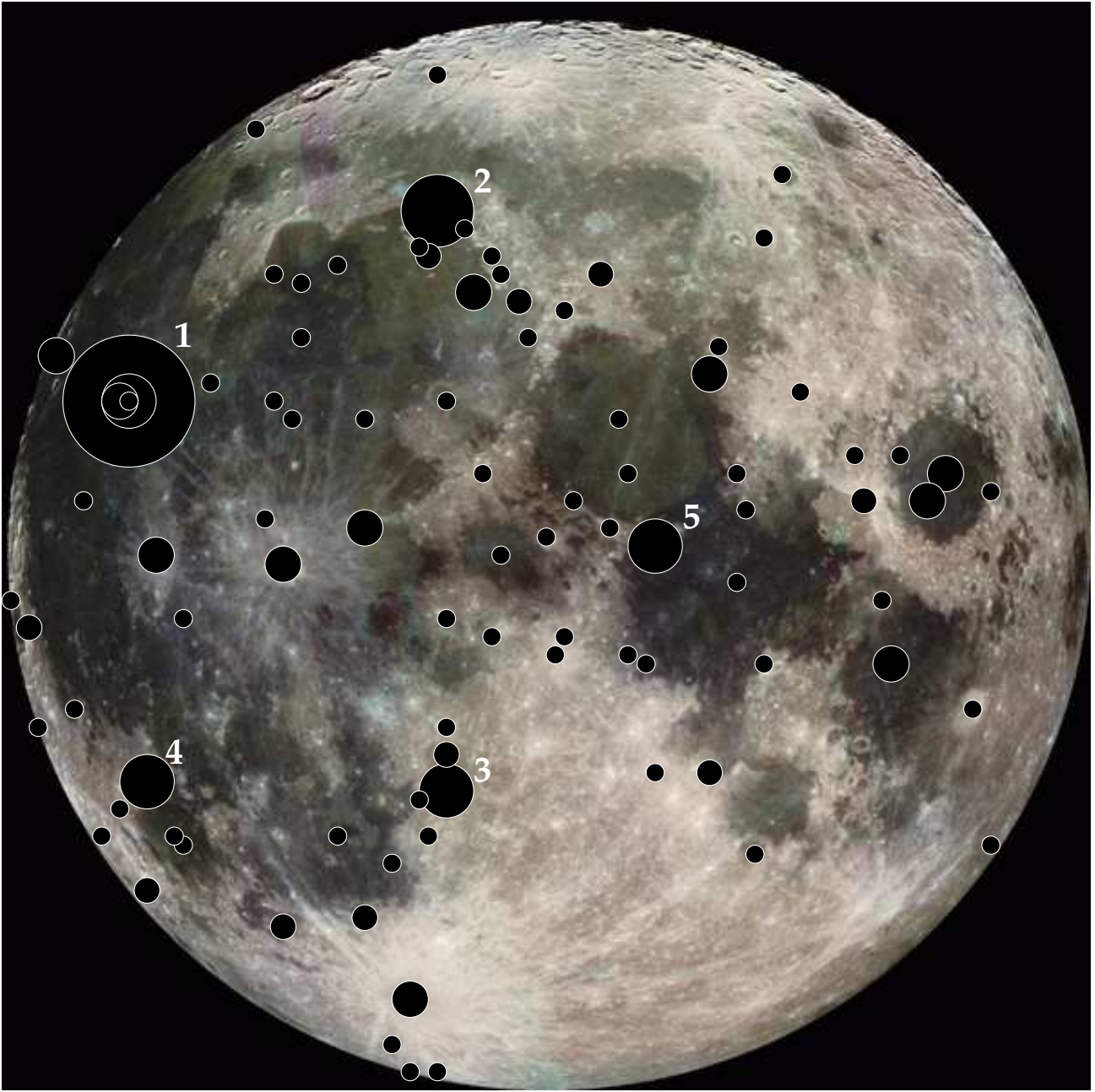}{5.9in}{-00}{090}{090}{-245}{-010}
\vskip 0.00in
\caption {
Distribution of TLP report loci as catalogued in Middlehurst et al.\ (1968),
with the exception of a minority of cases that are rejected for the reasons
detailed in the text.
The size of the symbols encodes the number of reports per features, as listed
in Table 1.
Note that the four symbols for features on the Aristarchus plateau overlap,
with the crater Aristarchus being the largest in the Figure.
Marked features include: 1) Aristarchus (enclosing Vallis Schr\"oteri, Cobra's
Head and Herotus, in decreasing order),
2) Plato,
3) Alphonsus
4) Gassendi
5) Ross D.~~~
(Photo credit: NASA)
}
\end{figure}

\begin{figure}
\plotfiddle{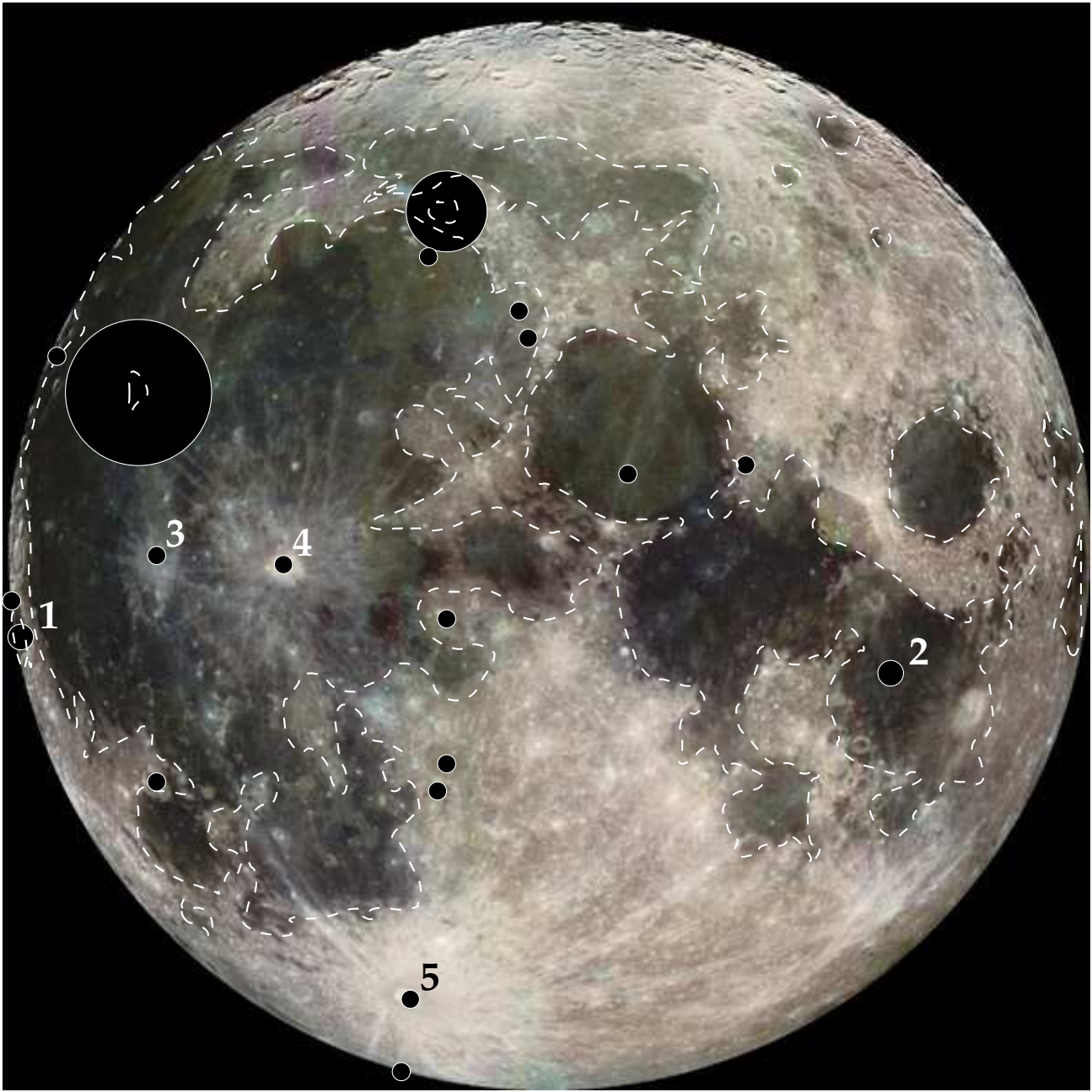}{6.0in}{-00}{090}{090}{-245}{-010}
\vskip 0.00in
\caption {
Distribution of TLP report loci that survive the robustness filter detailed in
the text, taken from those reports portrayed in Figure 1.
The size of the symbols encodes the number of reports per features, as listed
in Table 2.
Note that reports for the four features on the Aristarchus plateau overlap are
merged into a single symbol (the largest).
Marked features include several of note not marked in Figure 2:
1) Grimaldi,
2) Messier,
3) Kepler,
4) Copernicus,
5) Tycho.~~~
The dashed curves represent the locus of the mare/highland boundary adopted for
the calculations in the text.
(Photo credit: NASA)
}
\end{figure}

\begin{figure}
\plotfiddle{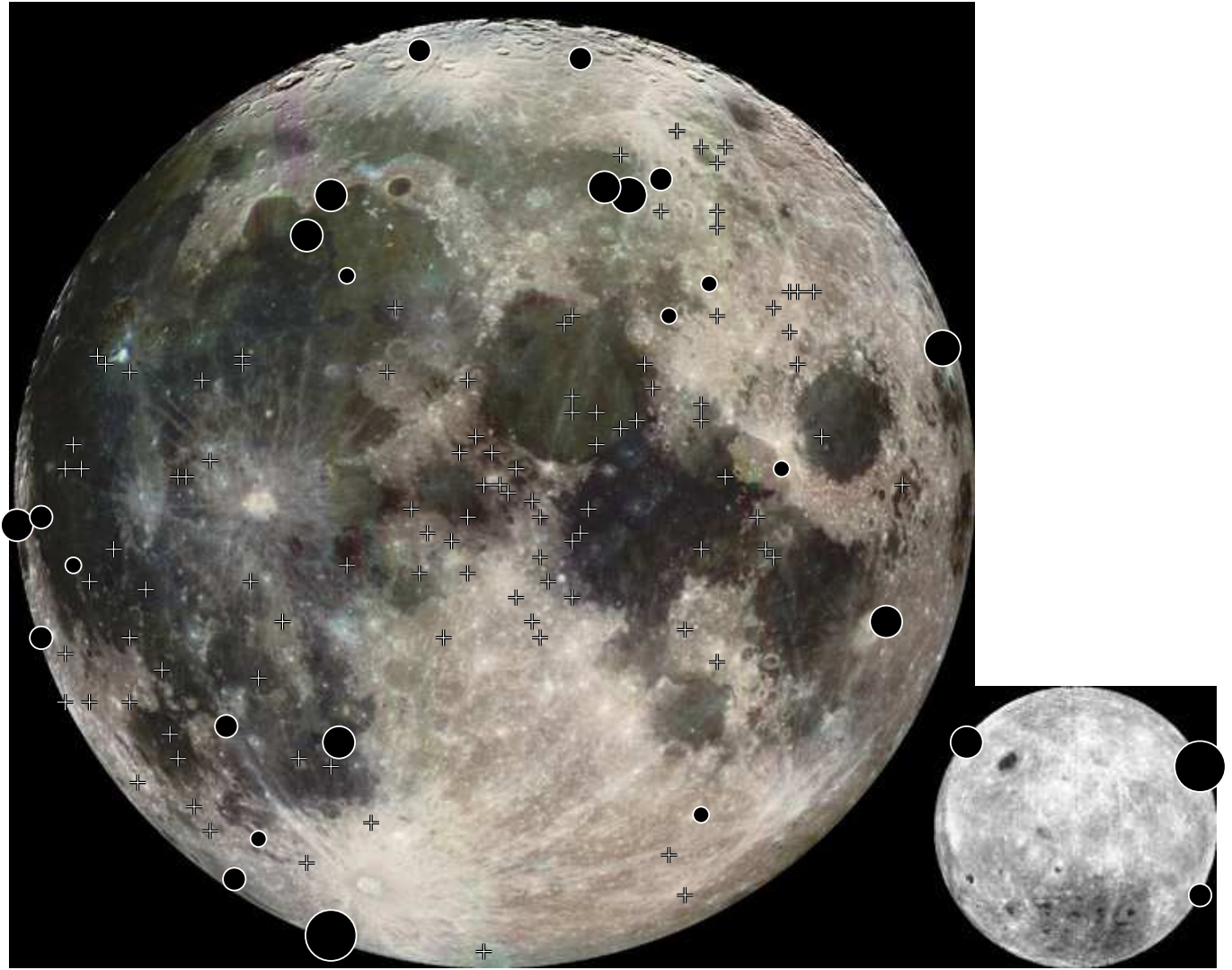}{6.7in}{-000}{075}{075}{-249}{+030}
\vskip 0.00in
\caption {
Distribution of shallow moonquakes from Nakamura et al.\ (1979) plotted with
with circular symbols indicating magnitude (from 0.7 to 3.2 on the HFT scale,
with magnitudes $\sim$1.0 less than their Richter magnitude --
Nakamura et al.\ 1979).
Note the complete absence of a signal from the Aristarchus plateau.
The three moonquake locations invisible from Earth are shown by similar symbols
on the smaller farside inset.
Most of these and the nearside quakes cluster near edges of maria.
The larger sample of deep moonquakes (Nakamura  2005, positions indicated by
crosses, typically accurate to $\sim 7^\circ$) is even more obviously
correlated with mare edges.
(Photo credit: NASA)
}
\end{figure}

\end{document}